\documentclass[8.5pt,twoside,twocolumn]{article}
\usepackage[super,sort&compress,comma]{natbib} 
\usepackage{times,mathptmx}
\usepackage{sectsty}
\usepackage{balance} 

\usepackage{graphicx} %eps figures can be used instead
\usepackage{lastpage}
\usepackage[format=plain,justification=raggedright,singlelinecheck=false,font=small,labelfont=bf,labelsep=space]{caption} 
\usepackage{fancyhdr}
\usepackage{amsmath}
\usepackage{graphicx}
\usepackage{color}
\usepackage{upgreek}
\usepackage[linkcolor = blue, citecolor = blue, urlcolor = blue, colorlinks = true]{hyperref}
\usepackage[version=3]{mhchem}
\usepackage{commath,amssymb}
\usepackage{bm}
\usepackage[capitalise]{cleveref}
\usepackage{textcomp}

\newcommand{\uM}{\, \mathrm{\upmu M}}

\newcommand{\ugperml}{\, \mathrm{\upmu g m l^{-1}}}
\newcommand{\mgperml}{\, \mathrm{mg m l^{-1}}}

\newcommand{\umpers}{\, \mathrm{\upmu m s^{-1}}}
\newcommand{\pers}{\, \mathrm{s^{-1}}}

\newcommand{\um}{\, \mathrm{\upmu m}}
\newcommand{\ul}{\, \mathrm{\upmu l}}

\begin{document}

\thispagestyle{plain}
\fancypagestyle{plain}{
\fancyhead[L]{\includegraphics[height=8pt]{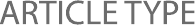}}
\fancyhead[C]{\hspace{-1cm}\includegraphics[height=20pt]{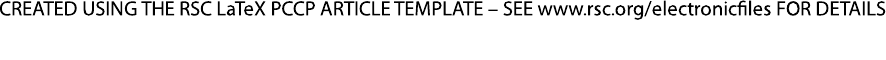}}
\fancyhead[R]{\includegraphics[height=10pt]{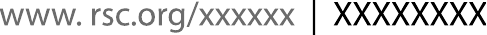}\vspace{-0.2cm}}
\renewcommand{\headrulewidth}{1pt}}
\renewcommand{\thefootnote}{\fnsymbol{footnote}}
\renewcommand\footnoterule{\vspace*{1pt}% 
\hrule width 3.4in height 0.4pt \vspace*{5pt}} 
\setcounter{secnumdepth}{5}

\makeatletter 
\def\subsubsection{\@startsection{subsubsection}{3}{10pt}{-1.25ex plus -1ex minus -.1ex}{0ex plus 0ex}{\normalsize\bf}} 
\def\paragraph{\@startsection{paragraph}{4}{10pt}{-1.25ex plus -1ex minus -.1ex}{0ex plus 0ex}{\normalsize\textit}} 
\renewcommand\@biblabel[1]{#1}            
\renewcommand\@makefntext[1]% 
{\noindent\makebox[0pt][r]{\@thefnmark\,}#1}
\makeatother 
\renewcommand{\figurename}{\small{Fig.}~}
\sectionfont{\large}
\subsectionfont{\normalsize} 

\newtoks\rowvectoks
\newcommand{\rowvec}[2]{%
 \rowvectoks={#2}\count255=#1\relax
 \advance\count255 by -1
 \rowvecnexta}
\newcommand{\rowvecnexta}{%
 \ifnum\count255>0
 \expandafter\rowvecnextb
 \else
 \begin{pmatrix}\the\rowvectoks\end{pmatrix}
 \fi}
\newcommand\rowvecnextb[1]{%
 \rowvectoks=\expandafter{\the\rowvectoks&#1}%
 \advance\count255 by -1
 \rowvecnexta}

\fancyfoot{}
\fancyfoot[LO,RE]{\vspace{-7pt}\includegraphics[height=9pt]{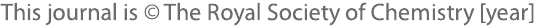}}
\fancyfoot[CO]{\vspace{-7.2pt}\hspace{12.2cm}\includegraphics{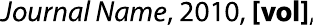}}
\fancyfoot[CE]{\vspace{-7.5pt}\hspace{-13.5cm}\includegraphics{RF}}
\fancyfoot[RO]{\footnotesize{\sffamily{1--\pageref{LastPage} ~\textbar  \hspace{2pt}\thepage}}}
\fancyfoot[LE]{\footnotesize{\sffamily{\thepage~\textbar\hspace{3.45cm} 1--\pageref{LastPage}}}}
\fancyhead{}
\renewcommand{\headrulewidth}{1pt} 
\renewcommand{\footrulewidth}{1pt}
\setlength{\arrayrulewidth}{1pt}
\setlength{\columnsep}{6.5mm}
\setlength\bibsep{1pt}

\twocolumn[
  \begin{@twocolumnfalse}
\noindent\LARGE{\textbf{Swimming in a Crystal$^\dag$}}
\vspace{0.6cm}

\noindent\large{\textbf{Aidan T. Brown,$^{\ast}$\textit{$^{a}$} Ioana D. Vladescu,\textit{$^{a}$}, Angela Dawson,\textit{$^{a}$} Teun Vissers,\textit{$^{a}$} Jana Schwarz-Linek,\textit{$^{a}$} Juho~S.~Lintuvuori,\textit{$^{a, b}$} and
Wilson C. K. Poon\textit{$^{a}$}}}\vspace{0.5cm}
%Please note that \ast indicates the corresponding author(s) but no footnote text is required. 

\noindent\textit{\small{\textbf{Received Xth XXXXXXXXXX 20XX, Accepted Xth XXXXXXXXX 20XX\newline
First published on the web Xth XXXXXXXXXX 200X}}}

\noindent \textbf{\small{DOI: 10.1039/b000000x}}
\vspace{0.6cm}
%Please do not change this text.

\noindent \normalsize{We study catalytic Janus swimmers and {\it Escherichia coli} bacteria swimming in a two-dimensional colloidal crystal. The Janus swimmers orbit individual colloids and hop between colloids stochastically, with a hopping rate that varies inversely with fuel (hydrogen peroxide) concentration. At high fuel concentration, these orbits are stable for 100s of revolutions, and the orbital speed oscillates periodically as a result of hydrodynamic, and possibly also phoretic, interactions between the swimmer and the six neighbouring colloids. Motile {\it E.~coli} bacteria behave very differently in the same colloidal crystal: their circular orbits on plain glass are rectified into long, straight runs, because the bacteria are unable to turn corners inside the crystal.}
\vspace{0.5cm}
 \end{@twocolumnfalse}
  ]

%Footnotes
\footnotetext{\dag~Electronic Supplementary Information (ESI) available online. See Appendix~\ref{videos} for captions to Supplementary Videos 1-4. See DOI: 10.1039/b000000x/.}

%Please use \dag to cite the ESI in the main text of the article.
%If you article does not have ESI please remove the the \dag symbol from the title and the above footnotetext.

\footnotetext{\textit{$^{a}$~SUPA, School of Physics and Astronomy, The University of Edinburgh, King's Buildings, Peter Guthrie Tait Road, Edinburgh, EH9 3FD, United Kingdom, E-mail: abrown20@staffmail.ed.ac.uk}}
\footnotetext{\textit{$^{b}$~Laboratoire de Physique des Solides, Universit\'e Paris-Sud, UMR 8502 - 91405 Orsay, France. }}

\section{Introduction}

Non-equilibrium systems pose a grand challenge cutting across many areas of physics. An exciting frontier concerns microscopic swimmers, both motile micro-organisms such as bacteria and algae, and, increasingly, a range of synthetic self-propelled colloids~\cite{poon2013physics, Howse}. Such swimmers attract interest in terms of their propulsive mechanisms~\cite{Howse, paxton04, gibbs09, howse07, ebbens12, brown14, ebbens14} and their collective behaviour~\cite{Joanny2014}. These aspects are coupled: the collective behaviour depends on how swimmers interact, and aspects of these interactions are directly related to the propulsive mechanism. 

Microswimmers interact with each other and with their surroundings via two classes of interactions. Various thermodynamic interactions (van der Waals, electrostatic, \ldots) are shared with passive particles. Their effect, as least on quiescent passive particles, can be treated using equilibrium statistical mechanics. A second class of interactions arises directly from self propulsion, and is responsible for the above-mentioned coupling between propulsive mechanism and collective behaviour. There is as yet no general theory to predict the effect of these detailed-balance-violating interactions. 

The most basic type of active interaction arises simply due to persistence, and explains the ubiquitous observation that confined microswimmers  accumulate at confining boundaries. Once a swimmer reaches a wall, it takes finite time to reorient and swim away~\cite{volpe11}. When many microswimmers are present, this interaction can induce phase separation~\cite{fily12, stenhammar13} and collective motion~\cite{dunkel13}. Another ubiquitous active interaction arises from the flow fields responsible for self propulsion around each microswimmer. Such  hydrodynamic interactions (HI) can produce qualitatively different behaviour from pesistence-induced interactions~\cite{Simha2004, Joanny2014}. Indeed, phase separation driven by the latter may be suppressed by HI~\cite{zottl14}. 

Considering persistence-induced interaction and HI often suffices for understanding biological microswimmers, because the bioenergetics powering self propulsion \cite{Jana2015} are typically internal. By contrast, synthetic active colloids often swim via self-generated external gradients, e.g. of electrostatic potential, chemical concentration or temperature. Phoretic interactions (PI)~\cite{soto14, uspal15, bickel13, ginot15, theurkauff12} arise from the coupling of these gradients to other surfaces and swimmers. These PI should in general have comparable magnitude to the HI, because the gradients are by definition strong enough to generate propulsion. 

Active interactions can be accessed experimentally by directly studying swimmer-swimmer interactions in dilute suspensions~\cite{liao07}, but also by looking at the interactions between swimmers and passive tracer particles~\cite{drescher10, drescher11, guasto10}, surfaces~\cite{chiang14, uspal15, ishimoto13, li09, li14, li14b} or porous media~\cite{takagi14, volpe11}. In this article, we focus on the interaction between swimmers and a model porous medium. We observe two popular swimmers, motile {\it Escherichia coli} bacteria and catalytic Pt-polystyrene Janus particles fuelled by \ce{H2O2}~\cite{howse07}, interacting with a close-packed 2D crystal of passive colloids. 

This environment has opposite effects on these two swimmers: it destroys the usual orbital motion of {\it E. coli} on plane surfaces~\cite{vigeant02, lauga06, shum10}, but induces orbital motion of the Janus swimmers around individual colloids in the crystal. We can explain the behaviour of {\it E. coli} cells simply, in terms of the steric hinderance between their long flagella bundles and their crystalline surroundings. Orbiting Janus swimmers are harder to understand. They hop stochastically between orbits round neighbouring colloids. The hopping rate scales inversely with the concentration of the fuel, \ce{H2O2}, but is independent of the speed of the swimmers at a particular fuel concentration, suggesting that orbital trapping is not just a hydrodynamic effect, but also involves PI. While there is insufficient information to pin down a specific trapping mechanism, we have characterised the trapping empirically in terms of a stiffness and an effective potential, which should help future theory construction in swimmer trapping at surfaces.

We observed extremely long-lived orbits at high \ce{H2O2} concentrations, lasting for 100s of rotations (10s of minutes). These could form the basis for useful devices -- most obviously, a microfluidic stirrer~\cite{takagi14}, and for studying the complex interactions between multiple orbiting swimmers.

Finally, we find that the orbital speed oscillates, resulting from interactions of an orbiting swimmer with the six nearest-neighbour colloids. These oscillations contain information on the propulsion mechanism, which dictates the character of the active interactions. However,  current, incomplete understanding of the phoretic propulsion of Janus swimmers~\cite{brown14, ebbens14} means that we cannot unambiguously disentangle the contributions of HI and PI to the observed speed oscillations. We therefore discuss our results in terms of a simple model swimmer interacting purely via HI with its surroundings. With future advances in understanding of PI in catalytic swimmers, our results should help constrain proposals for their propulsion mechanism. 

\begin{figure}[t]
\centering
\includegraphics[width=8.5 cm]{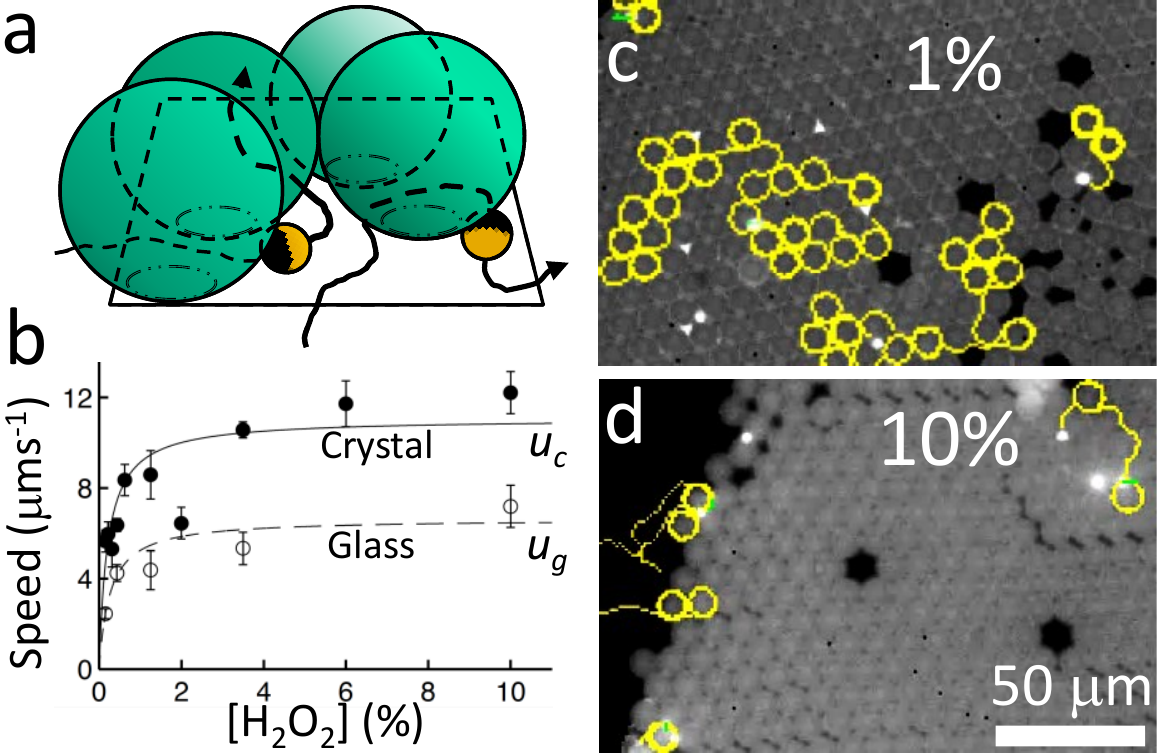}
\caption{a) Schematic of experimental setup. b) Mean ballistic speed $u$ inside ($u_c$, {\footnotesize $\bullet$}), and outside ($u_g$, $\circ$) crystal. Solid and dashed lines are fits to Eq.~\eqref{saturated speed}. c-d) Tracked video of Janus swimmers inside a colloidal crystal at c) 1$\%$ and d) 10$\%$ \ce{H2O2}, each of 3~min duration.}
\label{trackingJanus}
\end{figure}

\section{Materials and Methods}

\subsection{Janus Swimmers}
We prepared Janus particles (5 nm Pt sputtered on $2 \um$ diameter fluorescent polystyrene colloids from Invitrogen~\cite{brown14}) and suspended them at volume fraction $\sim 10^{-6}$ in aqueous \ce{H2O2} (Acros) solutions in chambers formed between glass slides (Menzel) and ${22\times22~{\rm mm}^2}$ glass coverslips (Bettering) with $300 \um$ thick parafilm spacers. On the coverslips, 2D colloidal crystals had been formed beforehand by depositing $10~\um$ diameter polystyrene colloids (Thermo-Fisher) at 1$\%$ v/v in water and evaporating at $70^\circ$C. The radii of the static colloids, ${R=5.06\pm 0.02~\um}$, and Janus swimmers, ${a=0.96\pm 0.04~\um}$, were determined by repeated (25$\times$) measurement of interparticle distances in close-packed 2D crystals. The electrophoretic mobilities of these colloids were obtained in $100~\uM$ \ce{NaNO3} (Fluka) using a Malvern Zetasizer. These were, for the ${2~\um}$ diameter, uncoated colloids, ${\mu_{\rm u}=(-5.3\pm0.1)\times 10^{-8}\mathrm{m^2V^{-1}s^{-1}}}$; for the ${10~\um}$, static colloids ${\mu_{\rm s}=(-4.4\pm0.5)\times 10^{-8}\mathrm{m^2V^{-1}s^{-1}}}$; and for the Janus particles, ${\mu_{\rm j}=(-5.3\pm0.1)\times 10^{-8}\mathrm{m^2V^{-1}s^{-1}}}$. Applying the Smoluchowski theory for the electrophoretic mobility, the surface charge density on these colloids is approximately $ q=\mu\eta\kappa$, with $\kappa^{-1}=30$ nm the Debye length, and $\eta=10^{-3}~{\rm Pa\;s}$ the viscosity of water, which gives ${ q_{\rm u}= q_{\rm j}=1.4\times 10^{-3}~{\rm Cm^{-2}}}$, and $ q_{\rm s}=1.2\times 10^{-3}~{\rm Cm^{-2}}$.

Our bottom-heavy Janus swimmers swim upwards~\cite{campbell13, brown14}. In addition, they become trapped and swim stably at any surface irrespective of its orientation~\cite{brown14}. Hence, to collect the swimmers in the colloidal crystal, we initially oriented the chamber with the coated coverslip uppermost, allowing the swimmers to collect there, before inverting the chamber for observation with an inverted epifluorescence microscope (Ti Eclipse, Nikon, $\times20$ objective) with a CCD camera (Eosens, Mikrotron). This inversion left swimmer behaviour unchanged. On colliding with a colloid at the edge of the crystal, the swimmers orbit that colloid in the wedge-like space between colloid and coverslip, Fig.~\ref{trackingJanus}a, before hopping out of the crystal or into orbit around another colloid. 

We varied the concentration of \ce{H2O2} between $0.1$ and $10\%$ v/v, and tracked the swimmers (using MATLAB~\cite{brown14,crocker96}) to determine the mean hopping rate and the swimming speed inside and outside the crystal at each \ce{H2O2} concentration~\cite{brown14}. We also measured at high magnification (using a 100$\times$ objective) the temporal variation in speed of swimmers in 10$\%$ \ce{H2O2}, with and without 100~$\uM$ \ce{NaNO3}, in orbit around colloids within and at the edge of colloidal crystals. From these videos, we extracted additionally the instantaneous orbital radius $\rho$, and the swimmer's orientation with respect to the tangent to the orbit in horizontal ($\beta$) and vertical ($\tau$) planes (defined in Fig.~\ref{oscillations}a). Further measurement details can be found in Appendix~\ref{app: geometry}.

\begin{figure}[t]
\centering
\includegraphics[width=8.5cm]{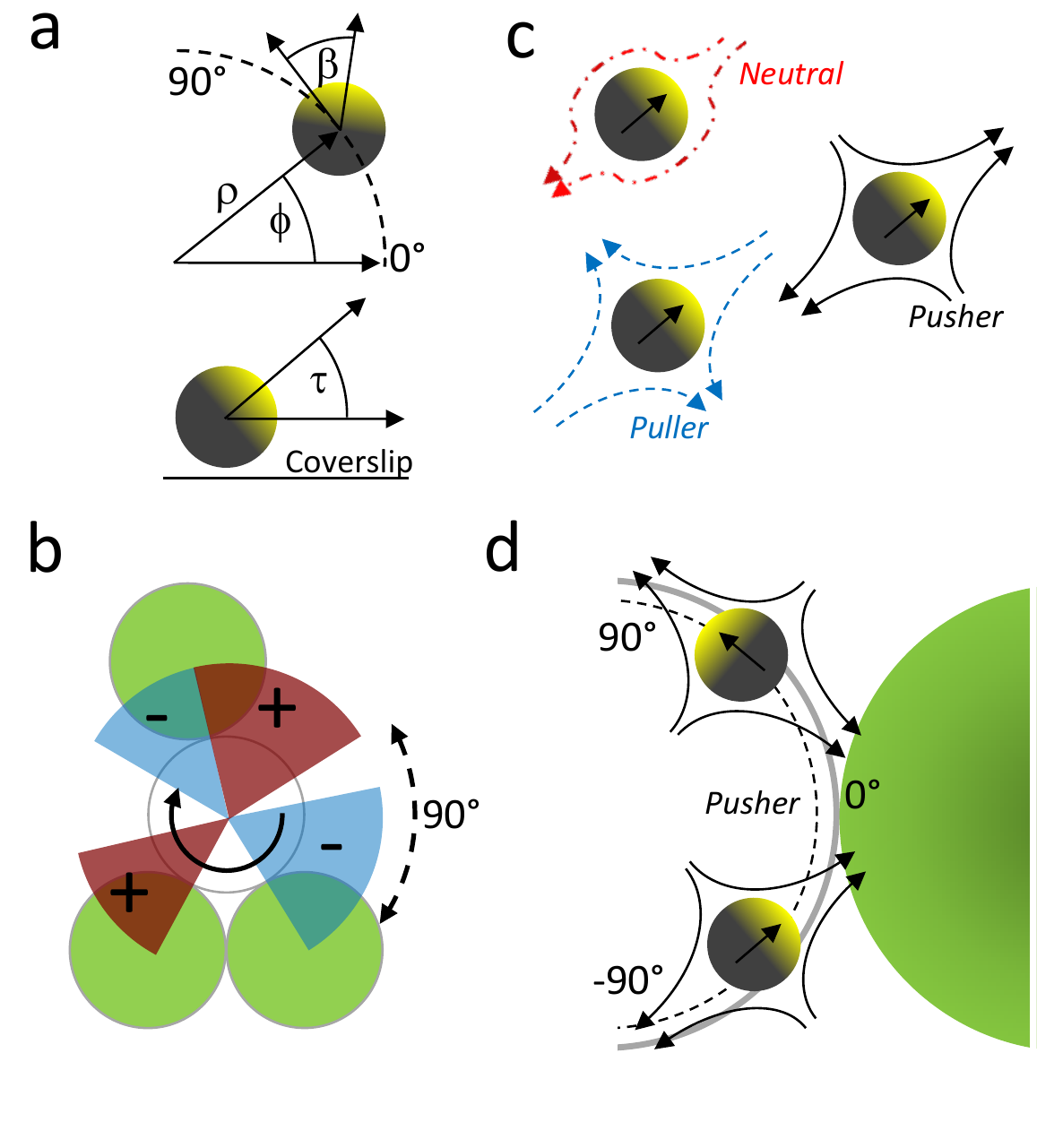}
\caption{a) Plan and side views of a Janus swimmer orbiting a colloid, showing the definition of the orientation angles $\phi$, $\beta$ and $\tau$, and the orbital radius $\rho$. b) Cartoon illustrating the reconstruction of the speed variation on passing near a single neighbouring colloid. For a clockwise orbit, points in the red regions are assigned $\phi>0$, and the blue regions $\phi<0$, with $\phi=0$ at the contact point between the central colloid and each respective neighbour. The uncoloured regions are not included, as they are within $90^\circ$ of two neighbouring colloids. c) Schematics of the flow-fields around neutral, pusher and puller swimmers, in the lab frame. c) Plan view of a pusher orbiting (dashed line) a central colloid (grey outline) with one neighbouring colloid (solid green).}
\label{oscillations}
\end{figure}

\subsection{{\it \textbf{E. coli}} bacteria}

Motile, GFP-labelled smooth-swimming {\it E.~coli} (strain AB1157 {\it $\Delta$cheY}), were cultured as previously described~\cite{Jana2015,vladescu14} and suspended in motility buffer (see Appendix~\ref{stain} for details) before being loaded into chambers with or without colloidal crystals. Further genetic modification permits tight binding of a fluorescent dye (Alexa Fluor 633) to the flagella bundle (see Appendix~\ref{stain} for details). The flagella-labelled GFP-expressing {\it E. coli} were observed in a Zeiss confocal microscope at 4 fps using laser excitation at 488 nm (for GFP) and 633 nm (for Alexa Fluor 633). We added 0.2 wt$\%$ TWEEN 20 to minimise adhesion of bacteria to the glass~\cite{Jana2015}.

\begin{figure}[]
\centering
\includegraphics[width=8.5 cm]{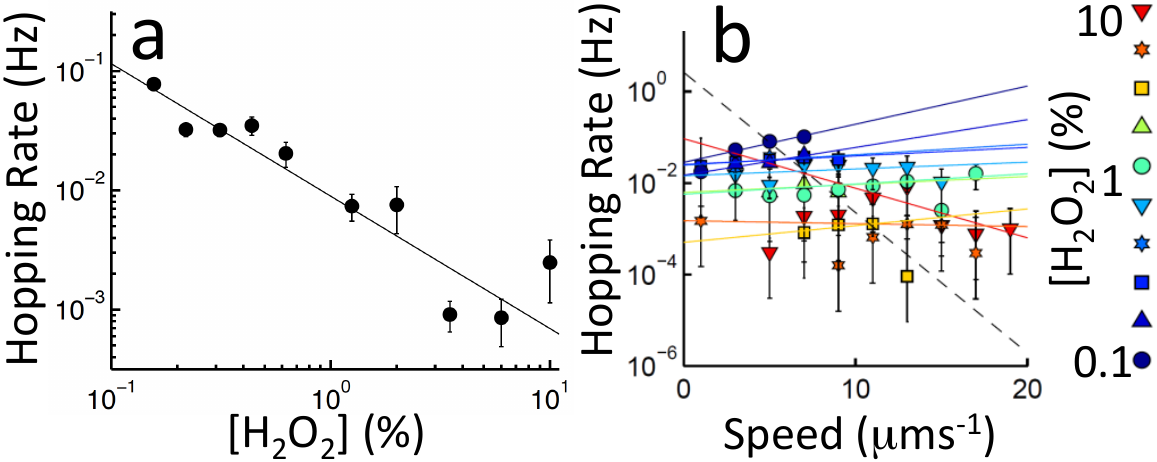}
\caption{a) Hopping rate vs.~\ce{H2O2} concentration, [\ce{H2O2}]. Solid line: $\Gamma \propto \mathrm{[\ce{H2O2}]^{-1}}$. b) Hopping rate binned by speed for each [\ce{H2O2}] (colour-coded). Solid lines: exponential fits within each [\ce{H2O2}]; dashed line: exponential fit through the mean of each data set.}
\label{MSDPlots}
\end{figure}

\section{Results}

\subsection{Orbital trapping of Janus swimmers}

Figure~\ref{trackingJanus}c-d shows typical trajectories of Janus swimmers in colloidal crystals in $1\%$ and $10\%$ \ce{H2O2}. In $1\%$ \ce{H2O2}, swimmers hop rapidly through the crystal (supplementary video SV1$^\dag$), whereas in 10$\%$ \ce{H2O2}, they remain in orbit around colloids near the edge of the crystal for many minutes. Figure~\ref{MSDPlots}a shows $\Gamma$, the rate at which swimmers hop between orbits, as a function of \ce{H2O2} concentration. The solid line is a fit to $\Gamma\propto[\ce{H2O2}]^{-1}$; a power law fit returns $\Gamma\propto[\ce{H2O2}]^{-1.06\pm0.03}$. 

\begin{figure}[]
\centering
\includegraphics[width=8.5 cm]{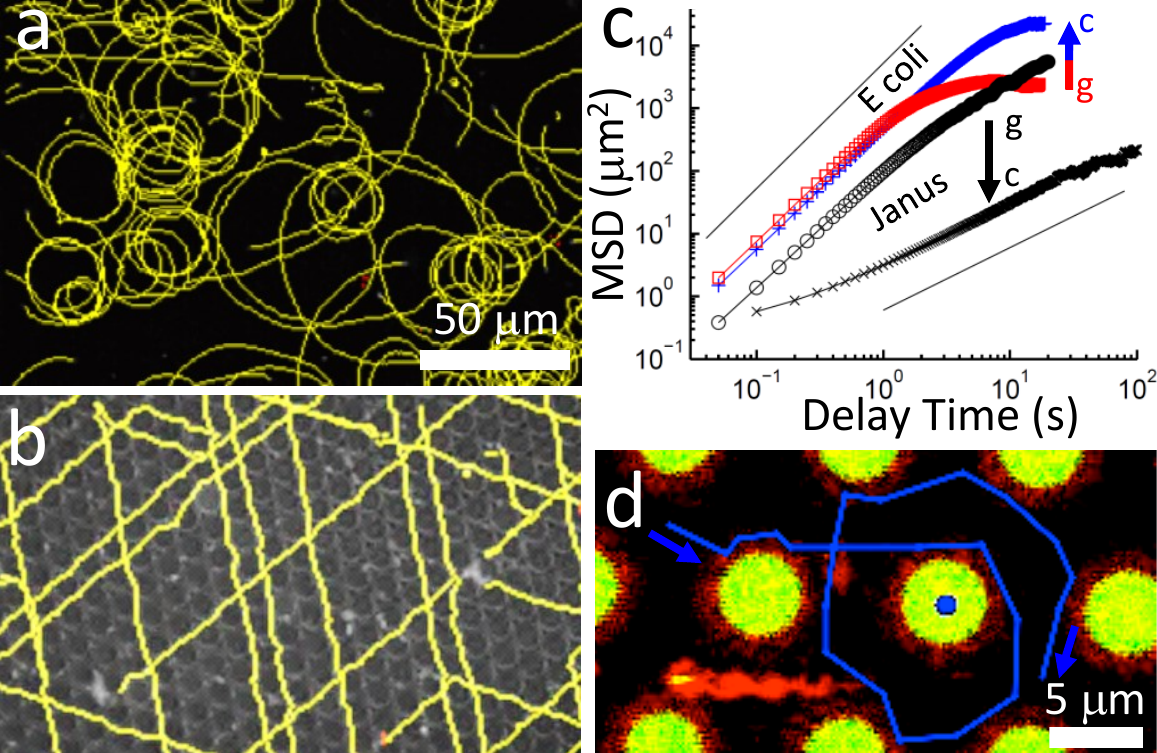}
\caption{Tracked videos of smooth-swimming {\it E.~coli} a) on plain glass, and b) inside a crystal. c) MSD on plain glass ($\circ$ = Janus; \textcolor{red}{$\square$} = {\it E.~coli}) and inside the crystal ($\times$ = Janus, \textcolor{blue}{+} = {\it E.~coli}). Solid lines: diffusive ($t$) and ballistic ($t^2$) scaling. Arrows highlight the effect of moving from glass into the crystal (g$\rightarrow$c). d)~Confocal image of a flagella stained (red) bacterium inside a colloidal crystal. Colloids (green) touch each other, but only a small, polar slice is visible. Blue: 6~s trajectory of a bacterium with shorter flagella (not shown).}
\label{trackingEcoli}
\end{figure}

Figure~\ref{trackingEcoli}c shows the mean squared displacement (MSD) of Janus swimmers inside and outside the crystal at $1\%$ \ce{H2O2}. Outside, the Janus swimmers move ballistically (MSD $\propto \Delta t^2$, with $\Delta t$ = delay time) at short times, becoming diffusive (${{\rm MSD}\propto\Delta t}$) at long times due to rotational diffusion~\cite{howse07}. Inside the crystal, they are diffusive at all times longer than the orbital time because orbiting destroys orientational memory. This distinction exists at all the \ce{H2O2} concentrations tested.

Varying \ce{H2O2} concentration affects the swimming speed, both inside and outside the crystal, as shown in Fig.~\ref{trackingJanus}b. At all \ce{H2O2} concentrations, the mean speed $\langle u_{\rm c}\rangle$ inside the crystal is larger than that on plain glass, $\langle u_{\rm g}\rangle$. The speed saturates at high \ce{H2O2} concentration, as previously observed~\cite{howse07}. This has been attributed to the saturation of Pt binding sites by \ce{H2O2} molecules, which gives a predicted speed of the form~\cite{howse07}
\begin{align}
 \langle u\rangle\,=\,\frac{u^*[\ce{H2O2}]}{[\ce{H2O2}]^*+[\ce{H2O2}]}\,, \label{saturated speed}
\end{align}
where $u^*$ is the saturation speed, and $[\ce{H2O2}]^*$ is the $\ce{H2O2}$ concentration at half maximum. The solid and dashed lines in Fig.~\ref{trackingJanus}b are best fits to Eq.~\ref{saturated speed} with $u_{\rm g}^*=6.6\pm 1~\umpers$ and $u_c^*=11.1\pm 2~\umpers $, and $[\ce{H2O2}]_{\rm c}^*=0.22\pm0.1\%$ and $[\ce{H2O2}]_{\rm g}^*=0.27\pm0.1\%$. 

On plain glass surfaces, the swimmers' motion also has a translational diffusive component, obtainable by fitting the MSD at short delay times~\cite{howse07, brown14, ebbens14} (the first five frames, i.e. 0.25 s), giving ${D_{\rm g}=0.21\pm 0.01~\um^2\pers}$ independent of \ce{H2O2} concentration, in agreement with previous measurement~\cite{howse07, brown14, ebbens14}.

At each \ce{H2O2} concentration, we measured a wide range of swimming speeds $u_{\rm c}$ of individual Janus particles, presumably due to variability of the Pt coating. Figure~\ref{MSDPlots}b shows the hopping rate $\Gamma$ as a function of individual swimming speed $u_{\rm c}$ for each \ce{H2O2} concentration data set. Each coloured line is an exponential fit to the data at a particular \ce{H2O2} concentration. The dashed black line is an exponential fit through the mean speed $\langle u_{\rm c}\rangle$ at each \ce{H2O2} concentration. The coloured lines are all much flatter than the mean fit, implying that there is little dependence of the hopping rate on swimming speed beyond the dependence on \ce{H2O2} concentration.

\subsection{Speed oscillations in Janus-swimmer orbits}

In 10$\%$ \ce{H2O2}, the orbits are extremely stable, and we tracked Janus swimmers orbiting single colloids within the crystal for 100s of revolutions (see SV2$^\dag$). The speed $u(\phi)$ as a function of the orbital angle $\phi$ (defined in Fig.~\ref{oscillations}a) shows sinusoidal oscillations, Fig.~\ref{edge oscillations}a). The solid curve is a fit of the form 
\begin{align}
u(\phi)\,=\,u_c\left\{1+\tilde{u}\cos{\left[6(\phi-\delta)\right]}\right\}\,, \label{speed oscillations}
\end{align}
with $\delta\in(-30^\circ, 30^\circ]$. The origin for $\phi$ is chosen so that the neighbouring colloids are at $\phi=0^\circ, 60^\circ$, etc. We measure from 17 videos a fractional amplitude $\tilde{u}=7.7\pm0.5\%$ and retardation $\delta=13.5\pm1.5^\circ$. Identical oscillations are found in $10\%$ \ce{H2O2} + $100~\uM$ \ce{NaNO3} with $\tilde{u}=7.3\pm 0.6\%$ and $\delta=14\pm 2^\circ$.

We also measured $u(\phi)$ for 35 swimmers orbiting colloids at the edge of the crystal with fewer (2 to 4) neighbours. For each swimmer, we analyse only those parts of its orbit where the swimmer does not have more than one neighbouring colloid within a $\pm 90^\circ$ sector, Fig.~\ref{oscillations}b. We then average these partial trajectories together to reconstruct the effect of a single neighbour. The $u(\phi)$ so constructed is shown in Fig.~\ref{edge oscillations}b.

Next, we determine the average position and orientation of the swimmer, with respect to the central colloid. To do so, we measure the angles $\beta$ and $\tau$ defining the orientation of the Janus particle relative to the orbital tangent, Fig.~\ref{oscillations}(a), and the average orbital radius, $\rho$, Fig.~\ref{geometry}. Following methods detailed in Appendix~\ref{app: geometry}, we find $\beta= 7\pm 2^\circ$, $\tau=1\pm2^\circ$ and $\rho=4.56 \pm 0.02\um$. These parameters do not fully constrain the 3D position of the Janus swimmer, but do place an upper bound on the distance $g_{\rm js}$ between the swimmer and the central colloid of $g_{\rm js} \lesssim 200$~nm. If we assume that $g_{\rm js}=g_{\rm jg}$, the distance between the swimmer and the glass surface, then $g_{\rm js}=g_{\rm jg}=70\pm10$ nm.

The mean temporal standard deviation of $\beta$ and of $\rho$ are $\sigma_\beta=1.9^\circ$ and $\sigma_\rho=12$~nm respectively. Note that these are averages of the standard deviations obtained from single orbits, rather than orbit-to-orbit variations. We found no oscillations in $\beta$ and $\rho$, so that $\sigma_\beta$ and $\sigma_\rho$ represent a combination of real temporal variabilites in these parameters and experimental uncertainties in their measurement.

Finally, we can determine a translational diffusion coefficient $D_{\rm c}$ for orbiting swimmers. Integrating Eq.~\eqref{speed oscillations} and assuming a diffusive term uncorrelated with the speed, we find, in the limit of time $t\rightarrow 0$, and $\tilde{u}\rightarrow 0$,
\begin{align}
\left\langle \left(\phi(t)-\phi(0)\right)^2\right\rangle\,=\,\left(1-\frac{\tilde{u}^2}{2}\right)\left(\frac{u_c}{\rho}\right)^2 t^2 + \frac{2D_{\rm c}}{\rho^2}t\,. \label{angular MSD}
\end{align}
Fitting Eq.~\ref{angular MSD} to data from the first five frames of 52 videos of swimmers orbiting in $10\%$ \ce{H2O2} gives $D_{\rm c}=0.082\pm 0.006~\um^2\pers$, independent of the swimming speed. To validate this procedure, we generated artificial data by stepping through the relevant Langevin equation 
\begin{align}
\phi(t+\Delta t)\,=\, \phi(t) + \frac{u_c\Delta t}{\rho}\left[1+\tilde{u}\sin\left(6\phi\right)\right]+\frac{\xi(t)}{\rho}\sqrt{D_{\rm c}\Delta t}\,, \label{timeStep}
\end{align}
for $10^5$ steps of $\Delta t=0.01~{\rm s}$ and with $\xi(t) = \pm 1$ at each step. Using experimental values for the other parameters, we recovered the input values for $D_{\rm c}$ and $u_{\rm c}$ by fitting to Eq.~\eqref{angular MSD} and averaging over 100 simulated swimmers.

\subsection{Rectification of {\it \textbf{E. coli}} trajectories}

Fig.~\ref{trackingEcoli}a-b shows trajectories of {\it E. coli} bacteria outside and inside a colloidal crystal. On plain glass, these bacteria circulate clockwise (viewed from the fluid side) due to their rotating flagella~\cite{vigeant02, lauga06, shum10}. The crystal rectifies this circulation into straight trajectories. Figure~\ref{trackingEcoli}c show the MSD of {\it E. coli} averaged over 5 videos per curve. Outside the crystal, trajectories are initially ballistic, levelling off at long times due to the circular motion. Inside the crystal, the ballistic regime is extended, because the bacteria cannot circulate.

We also observed individual bacteria with stained flagella in more detail. As shown in Fig.~\ref{trackingEcoli}d, and SV3$^\dag$, a few cells with shorter flagella ($\sim 3~\um$, compared to $\sim 7~\um$ on average) do occasionally show circular trajectories. These circular trajectories have radii at the lower end of the range of orbital radii observed on plain glass, so that the circulation is probably just that produced by the bacteria themselves, and is not induced by the colloids. 

\section{Discussion}

\subsection{Orbital trapping of Janus swimmers}

We first discuss the orbital trapping of the Janus swimmers in terms of its ubiquity and relationship with trapping at surfaces and edges, its stability, and potential trapping mechanisms.

\subsubsection{Ubiquity\;}

As previously noted~\cite{brown14}, micron-sized catalytic Janus swimmers are stably trapped at glass surfaces. Additionally, we have observed their trapping on the surfaces of $100~\um$ polystyrene beads (Thermo Scientific), and of hexadecane (Sigma Aldrich) droplets. If such trapping is general, then the orbital behaviour found in this work, Fig.~\ref{trackingJanus}c,d, should also be generic. Indeed, we have also observed stable orbits around silica beads (Bangs Labs), hexadecane droplets, and oxygen bubbles from \ce{H2O2} decomposition. These swimmers also follow the internal edge of water droplets on glass in air or in oil, and orbit around the horizontal axis between two colloids within the crystals studied here when a defect in the structure leaves sufficient space to do this~\footnote{In a perfectly hexagonally ordered layer of 10$~\um$ diameter spheres, the interstice between three neighbouring spheres is too small for the passage of a 2$~\um$ diameter sphere.}. Finally, orbiting behaviour was previously reported for Pt-Au nanorods~\cite{takagi14}.

\subsubsection{Effective trapping potential \;}
Our stable orbit corresponds to a fixed point in a 4-dimensional phase space (two orientational and two translational degrees of freedom, assuming that the swimmer is axisymmetric and that we can ignore the interactions with neighbouring colloids). Assuming that there are no limit cycles near this fixed point, we can treat the swimmer as though it were trapped in a potential well in $\beta,~\rho$ space and use the equipartition theorem to translate the measured standard deviations, $\sigma_\rho$ and $\sigma_\beta$ into the stiffness of the trapping potential, $k_\rho=k_BT/\sigma_\rho^2=3\times 10^{-5}~{\rm Jm^{-2}}$ and $k_\beta=k_BT/\sigma_\beta^2=4\times 10^{-18}~{\rm J}$ in these two directions. Similarly, from the hopping rate, $\Gamma$, we can estimate the depth of the effective trapping potential $U$ using the Kramers theory of barrier escape, which gives an escape frequency of~\cite{hanggi90}
\begin{align}
\Gamma\,\approx \,A\exp{\left(-\frac{U}{k_BT}\right)}\, .\label{effective potential}
\end{align}
The attempt rate $A$ depends on the form of the potential, which is unknown; but $A$ typically has the form ~\cite{hanggi90}
\begin{align}
A\,=\,\frac{k D}{2\pi k_BT}\,,
\end{align}
where $k$ has the dimensions of stiffness and $D$ is a relevant diffusivity. In our case, the simplest escape routes come from large fluctuations in $\beta$ or $\rho$. For fluctuations in $\rho$, we estimate $k=k_\rho$. To estimate the relevant $D$, we have to know the effect of nearby surfaces on diffusion normal to these surfaces. It is known that for a surface-to-surface gap of $g = 0.1a$, the diffusivity $D_\rho\sim$ is 10\% of the free-particle diffusivity for a swimmer close to a single plane wall~\cite{cox67}. Using these values, we find $A_\rho\sim30~\pers$, which, together with $\Gamma=10^{-3}~\pers$ at $10\%$ \ce{H2O2},  gives $U\sim12k_BT$ from Eq.~\eqref{effective potential}. Considering fluctuations in  $\beta$ gives a similar result. Because the measured $\sigma_\rho$ and $\sigma_\beta$ are upper bounds on the true standard deviations, our estimate for $U$ is an approximate lower bound.

\subsubsection{Phase stability\;}

The low measured diffusivity in the angle $\phi$, Eq.~\ref{angular MSD}, means that these orbiters show long-term phase stability. The number of revolutions over which phase is retained, or the quality factor, $Q$, is related to the time taken to diffuse $\pi/2$ away from the ballistic prediction $\phi=u_c t/\rho$:
\begin{align}
Q\,=\,\frac{\pi u_c\rho}{16D_{\rm c}}\,.
\end{align}
Our data give $Q=107\pm 5$. Such phase stability means that it would be interesting to study the coupling between two neighbouring orbiters, particularly to compare with previous work on beads dragged in circular orbits by optical traps~\cite{kotar13}.

\subsubsection{Passive interactions and persistence\;} The distances between the swimmer and the colloid and glass surfaces are of order $100~{\rm nm}$. This allows us to exclude almost all passive interactions from being significant for orbital trapping. Dispersion forces simply are too short in range (order nm). While electrostatic interactions have sufficient range, all the passive, electrostatic interactions should be repulsive, since all relevant surfaces are negatively charged under our conditions~\cite{gu00, lameiras08}. The effect of gravity can also be ruled out, since these swimmers become trapped on both the upper and lower surfaces of glass slides~\cite{brown14}, and on vertical, glass surfaces~\cite{campbell13}, and orbit within colloidal crystals on all of these surfaces.

We can also rule out the possibility that the swimmers are trapped in stable orbits simply because they do not rotate away quickly enough, as has been observed for other catalytic swimmers trapped at plane surfaces~\cite{volpe11}. The trapping timescales at surfaces~\cite{brown14} and within the crystal are much longer than the rotational diffusion time, which is $\sim 5~{\rm s}$ for these swimmers~\cite{brown14}. In addition, unlike at a plane surface, such a mechanism could not trap a swimmer in orbit without some additional orientational constraint. Another possibility is that the swimmers are trapped just by an orientational constraint, i.e. that they orient towards the surface, and their propulsion maintains them there. However, our swimmers are oriented away from the colloid surface, so this mechanism cannot apply here.

\subsubsection{HI and PI\;}
It has been suggested~\cite{takagi14} that the orbital trapping of Pt-Au nanorods is purely hydrodynamic. In trapping by pure HI, the hopping rate $\Gamma$ would be determined by a balance between HI, which maintain a stable swimmer orientation and position, and thermal fluctuations, which disrupt this stability~\cite{takagi14}. HI increase with swimming speed, which would give a strong negative correlation between swimming speed and $\Gamma$. 

Our measured $\Gamma$ shows no such direct speed dependence. At each \ce{H2O2} concentration, there is a wide variation in $u_{\rm c}$, but there is no systematic variation of $\Gamma$ with $u_{\rm c}$ (Fig.~\ref{MSDPlots}b). Hence, the trapping is strongly  dependent on \ce{H2O2} concentration (Fig.~\ref{MSDPlots}a), but via some speed-independent mechanism. This indicates a significant contribution from PI, as previously discussed theoretically for self-diffusiophoresis on plane surfaces~\cite{uspal15}. While data do not unambiguously pin down a specific mechanism, the observed $\Gamma\propto[\ce{H2O2}]^{-1}$ dependence provides a strong constraint for future theories on the propulsion and interaction of these swimmers.

\begin{figure}[ht!]
\centering
\includegraphics[width=8.5cm]{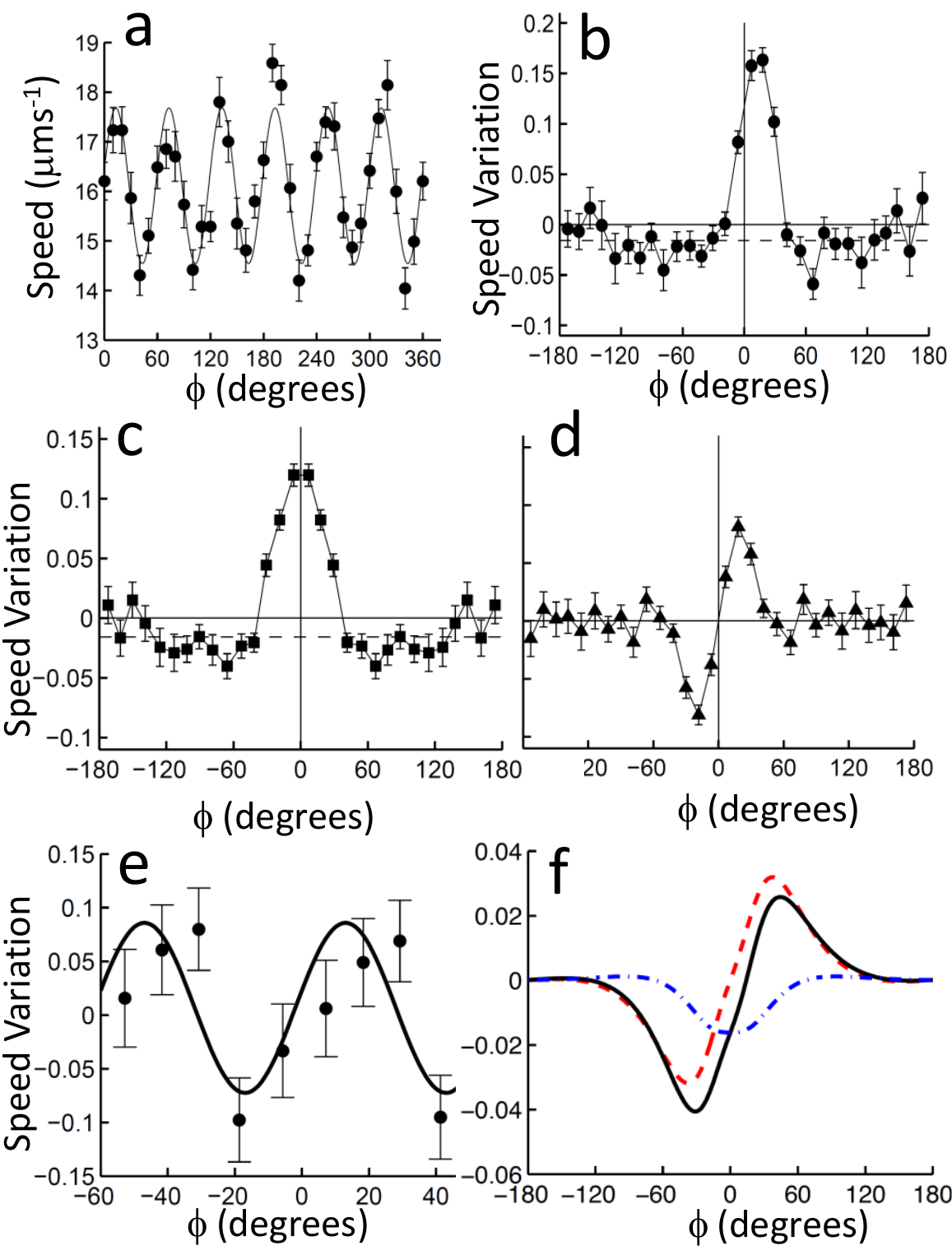}
\caption{a) Typical oscillations in orbital speed, from $\sim$200 revolutions of a Janus swimmer inside a crystal, in 10$\%$ \ce{H2O2}. Solid curve: 60$^\circ$ Fourier component.  b) Reconstructed, fractional speed variation caused by orbiting past a single neighbouring colloid at $\phi=0^\circ$, averaged over 35 swimmers. c) Symmetric, and d) antisymmetric parts of the speed variation in b. Dashed, horizontal lines indicate the average speed outside the perturbed region, $-90^\circ<\phi<90^\circ$. e) Periodic oscillation generated by repeatedly shifting the raw data in b by $60^\circ$ and adding the resulting speed variations together. The solid line is the $60^\circ$ sinusoidal components. f) Theoretical fractional speed variation (black, solid) of a pusher with stresslet amplitude $\alpha=1$ swimming past a neighbouring colloid, as described in the text. The antisymmetric (red, dashed) and symmetric (blue, dot-dashed) components are also shown.}
\label{edge oscillations}
\end{figure}

\subsection{Speed oscillations in Janus swimmer orbits}

We next consider the possible origins of the oscillations in orbital speed of Janus swimmers, which clearly arise from interactions between the swimmer and the neighbours to the colloid being orbited. 

\subsubsection{Passive interactions\;} The surface to surface distance between the Janus swimmer and these neighbouring colloids is at least $800$~nm, so, again, the only plausible passive interaction mechanism is electrostatic. Adding $100~\uM$ \ce{NaNO3} (with a Debye length, $\kappa^{-1} \lesssim 30~$nm) left the oscillations unchanged, and in this case we can estimate the minimum charge density $q$ needed to give the observed oscillation amplitude. The screened potential between two charged spheres is~\cite{leunissen05}
\begin{eqnarray}
U\,=\,\frac{4\pi q^2 aR}{\epsilon\kappa^2(g+a+R)}\exp{(-\kappa d)}\,, \label{Coulomb Potential}
\end{eqnarray}
where $g$ is the surface-surface separation and $\epsilon$ is the dielectric permittivity of water. The maximum amplitude is therefore 
\begin{eqnarray}
\delta u_{\rm max}\,\approx \,\frac{1}{6\pi\eta a}\left|\frac{\partial U}{\partial g}\right|\,.\label{max speed}
\end{eqnarray}
The observed amplitude of $\delta u\sim1~\umpers$ requires an unrealistically high charge density of $ q \gtrsim 1~\mathrm{Cm^{-2}}$, approximately 1000 times our measured values of $10^{-3}~{\rm Cm^{-2}}$. Hence, passive electrostatic interactions cannot generate the observed speed oscillations.

\subsubsection{Fuel Concentration Variation\;} In principle, the oscillations could be due to spatial variations in absolute \ce{H2O2} concentration directly affecting the swimming speed. However, the propulsion speed is insensitive to the concentration at $10\%$ \ce{H2O2} (Fig.~\ref{trackingJanus}b), and the measured \ce{H2O2} consumption rate~\cite{brown14} implies that these swimmers will deplete the \ce{H2O2} concentration at their surface by only around $1\%$ of the bulk concentration. Therefore, these local variations in fuel concentration cannot account for the observed speed oscillations.

\subsubsection{HI and PI\;} To examine the potential roles of HI and PI in generating the observed speed oscillations, we examined the orbits of swimmers orbiting colloids at the edge of our crystals with fewer (2-4) neighbours, and reconstructed the effect of a single neighbour on the orbiting speed (Fig.~\ref{edge oscillations}b). A single peak is seen just after the swimmer passes a neighbour (Fig.~\ref{edge oscillations}b), consistent with the positive retardation seen inside the crystal. The speed variation is practically nil by $\phi=60^\circ$, i.e., a swimmer is little affected by colloids beyond the one or two neighbouring colloids which it is closest to at any point. 

We now sum six suitably shifted copies of the data in Fig.~\ref{edge oscillations}b to predict the speed oscillations expected for a swimmer orbiting a colloid in the bulk of our colloidal crystal (Fig.~\ref{edge oscillations}e, points). The $60^\circ$-period sinusoidal component (Fig.~\ref{edge oscillations}e, line) has amplitude $7\%$ and retardation $13^\circ$, consistent with experiments performed inside the crystal.  

Interpreting these oscillations is difficult due to uncertainties in the swimmers' propulsion mechanism. While there is strong evidence against the originally-proposed self-diffusiophoretic mechanism~\cite{brown14}, the details of the true mechanism, which appears to be some version of self-electrophoresis, remain obscure~\cite{brown14, ebbens14}. Moreover, the electric field produced by a self-electrophoretic swimmer could produce PI in two ways: by being reflected from dielectric discontinuities, here liquid--static-colloid interfaces; and by advection in the electroosmotic flow induced by the interaction of the field with the fixed charge on those interfaces. Unlike passive electrostatic interactions, these active electrostatic interactions are long ranged: they involve ionic currents, so are not subject to the equilibrium ionic screening. 

Due to these complications, we neglect PI here and discuss speed variations expected for a model swimmer in our geometry subject only to HI, and compare these to experiments. Understanding the role of hydrodynamics should, of course, contribute to a future full theory that also involves PI. 

We use the lowest-order model for the flow field $\mathbf{u}$ around a force-free swimmer, which is a stresslet (force dipole) field $\mathbf{S}$ of amplitude $\alpha$. So, $\mathbf{u}=\alpha\mathbf{S}$ with, in spherical polars,
\begin{align}
\mathbf{S}\,=\,\frac{1}{2r^2}\left(1+3\cos2\theta\right)\hat{\mathbf{r}}\,,
\end{align}
where $\theta$ is the angle from the swimmer's unperturbed propulsion direction and $r\hat{\mathbf{r}}$ is the vector displacement from its center. Different types of swimmers are distinguished by the symmetry of their surrounding flow fields, i.e. the sign of the prefactor $\alpha$ (Fig.~\ref{oscillations}c). Pushers (e.g. {\it E. coli}) have amplitude $\alpha>0$, while pullers (e.g. various {\it Chlamydomonas} algae) have $\alpha<0$. In neutral swimmers, $\alpha=0$, and higher order terms become important.

In Fig~\ref{edge oscillations}c-d, we split the observed speed variation up into antisymmetric and symmetric parts around the point where a swimmer passes a neighbouring colloid. Comparison of Fig.~\ref{oscillations}c-d with Fig.~\ref{edge oscillations}d shows our observations are consistent with a pusher.  As a pusher approaches the neighbouring colloid, the fluid pushed out in front is reflected by the colloid and slows the swimmer down; once the swimmer has passed the colloid, the fluid pushed out behind speeds it up. 

If this speed variation is solely due to the stresslet component, we can estimate $\alpha$ by calculating the approximate hydrodynamic interaction between a free stresslet swimmer in circular orbit and a spherical surface outside that orbit. We use the measured position (assuming $g_{\rm js}=g_{\rm jg}=70$ nm) and orientation ($\tau, \beta$) of the swimmer with respect to the central colloid, and a far-field analytical expression for the interaction of a stresslet with a spherical surface~\cite{spagnolie15}. Calculational details are in Appendix~\ref{oscillations}. Figure~\ref{edge oscillations}e shows the predicted fractional speed variation $\Delta u/u_0$ for $\alpha=u_0 a^2$ (black solid), and the antisymmetric (red dashed) and symmetric (blue dot-dashed) components of this variation. The predicted speed variation would be completely antisymmetric for a swimmer oriented tangent to the orbit, but the inclination of the swimmer away from the orbit breaks this symmetry slightly. To match the peak heights between Fig.~\ref{edge oscillations}d and the antisymmetric component of Fig.~\ref{edge oscillations}f requires $\alpha\sim30~\um^3\pers$. This appears reasonable, as ${\it E. coli}$ of a similar size moving at a similar speed have a measured $\alpha=40~\um^3\pers$~\cite{drescher11}. Nevertheless, our value is no more than a very rough estimate. Moreover, the stresslet contribution clearly cannot fully explain the observed speed oscillation. Further analysis of the HI, including interaction with the central colloid and the plane surface, as well as future advances enabling the including of PI, will be necessary before firm conclusions can be drawn. 

\subsection{Rectification of {\it \textbf{E. coli}} trajectories}

The behaviour of {\it E. coli} can be explained more simply. Their typical circulation radius (Fig.~\ref{trackingEcoli}a) is much larger than the inter-colloid spacing, and at $\sim 7~\um$, their flagella are likely to hinder turning out of the straight channels between colloids. Occasionally, bacteria do briefly orbit individual colloids, but imaging {\it E. coli} with fluorescent flagella, shows that these cells  typically have shorter, $\sim3~\um$ flagella (Fig.~\ref{trackingEcoli}d and SV3$^\dag$), and so should also have a naturally tighter circulation radius than bacteria with longer flagella~\cite{lauga06}. Unlike Janus swimmers, bacteria do not appear to be trapped by the colloid at the centre of their orbit, and do not approach it closely (Fig.~\ref{trackingEcoli}d). 

It is interesting that the complex environment of the colloidal crystal can effectively simplify the trajectories of {\it E. coli} bacteria compared to their behaviour on plane surfaces. This may have applications in studying various -taxes (chemotaxis, phototaxis etc.) on surfaces, where circulation would normally prevent the bacteria from biassing their motion along favourable gradients. 

\section{Conclusion}

We have studied the behaviour of catalytic Janus swimmers and motile {\it E. coli} bacteria inside a model 2D colloidal crystal. The effect of this porous environment on these two swimmers is, respectively, to create and destroy, orbital motion. 

Our measurement of the behaviour of Janus swimmers inside the colloid crystal has generated a wealth of data on their behaviour in this environment, including detailed characterisation of orbital speed oscillations. These data set constraints for future work on the propulsion mechanism of these swimmers. Such understanding would then allow an assessment of the importance of PI in our crystalline geometry. If PI turn out to be minor, then our analysis of HI suggests that Janus swimmers are pushers with similar dipolar flow field amplitude to {\it E. coli}. In that case, then the very different response to the crystalline environment of these two self-propelled particle systems is noteworthy: many theoretical calculations and simulations assume, at least implicitly, that it is fruitful to discuss `generic pusher behaviour'. Our data suggest otherwise.  

Our observations immediately suggest other studies. For example, the circulation of {\it E. coli} next to surfaces presents an obstacle to the study of chemotaxis, which crystalline rectification would presumably overcome. The stable Janus swimmer orbits  at high fuel concentration could form the basis for constructing various microfluidic devices, e.g., a mixer on the micro level~\cite{takagi14}. \\
~\\
{\it Acknowledgements} - This work was funded by UK EPSRC grant EP/J007404/1, EU Intra-European Fellowships 623364 LivPaC FP7-PEOPLE-2013-IEF and 623637 DyCoCoS FP7-PEOPLE-2013-IEF, and ERC Advanced Grant ERC-2013-AdG 340877-PHYSAPS. We thank Mike Cates, Elise Darmon, Davide Marenduzzo, Elliot Marsden, Alexander Morozov, Anne Pawsey, Jerko Rosko and Joakim Stenhammar for helpful discussions.

{\footnotesize
\bibliography{REFERENCELIST} %your .bib file

\providecommand*{\mcitethebibliography}{\thebibliography}
\csname @ifundefined\endcsname{endmcitethebibliography}
{\let\endmcitethebibliography\endthebibliography}{}
\begin{mcitethebibliography}{49}
\providecommand*{\natexlab}[1]{#1}
\providecommand*{\mciteSetBstSublistMode}[1]{}
\providecommand*{\mciteSetBstMaxWidthForm}[2]{}
\providecommand*{\mciteBstWouldAddEndPuncttrue}
  {\def\EndOfBibitem{\unskip.}}
\providecommand*{\mciteBstWouldAddEndPunctfalse}
  {\let\EndOfBibitem\relax}
\providecommand*{\mciteSetBstMidEndSepPunct}[3]{}
\providecommand*{\mciteSetBstSublistLabelBeginEnd}[3]{}
\providecommand*{\EndOfBibitem}{}
\mciteSetBstSublistMode{f}
\mciteSetBstMaxWidthForm{subitem}
{(\emph{\alph{mcitesubitemcount}})}
\mciteSetBstSublistLabelBeginEnd{\mcitemaxwidthsubitemform\space}
{\relax}{\relax}

\bibitem[Poon(2013)]{poon2013physics}
W.~C.~K. Poon, \emph{Physics of Complex Colloids}, Societ\`a Italiana di
  Fisica, Bologna, 2013, pp. 317--386 (= arXiv:1306.4799)\relax
\mciteBstWouldAddEndPuncttrue
\mciteSetBstMidEndSepPunct{\mcitedefaultmidpunct}
{\mcitedefaultendpunct}{\mcitedefaultseppunct}\relax
\EndOfBibitem
\bibitem[Ebbens and Howse(2010)]{Howse}
S.~J. Ebbens and J.~R. Howse, \emph{Soft Matter}, 2010, \textbf{6},
  726--738\relax
\mciteBstWouldAddEndPuncttrue
\mciteSetBstMidEndSepPunct{\mcitedefaultmidpunct}
{\mcitedefaultendpunct}{\mcitedefaultseppunct}\relax
\EndOfBibitem
\bibitem[Paxton \emph{et~al.}(2004)Paxton, Kistler, Olmeda, Sen, St.~Angelo,
  Cao, Mallouk, Lammert, and Crespi]{paxton04}
W.~F. Paxton, K.~C. Kistler, C.~C. Olmeda, A.~Sen, S.~K. St.~Angelo, Y.~Cao,
  T.~E. Mallouk, P.~E. Lammert and V.~H. Crespi, \emph{J. Am. Chem. Soc.},
  2004, \textbf{126}, 13424\relax
\mciteBstWouldAddEndPuncttrue
\mciteSetBstMidEndSepPunct{\mcitedefaultmidpunct}
{\mcitedefaultendpunct}{\mcitedefaultseppunct}\relax
\EndOfBibitem
\bibitem[Gibbs and Zhao(2009)]{gibbs09}
J.~G. Gibbs and Y.-P. Zhao, \emph{Appl. Phys. Lett.}, 2009, \textbf{94},
  163104\relax
\mciteBstWouldAddEndPuncttrue
\mciteSetBstMidEndSepPunct{\mcitedefaultmidpunct}
{\mcitedefaultendpunct}{\mcitedefaultseppunct}\relax
\EndOfBibitem
\bibitem[Howse \emph{et~al.}(2007)Howse, Jones, Ryan, Gough, Vafabakhsh, and
  Golestanian]{howse07}
J.~R. Howse, R.~A.~L. Jones, A.~J. Ryan, T.~Gough, R.~Vafabakhsh and
  R.~Golestanian, \emph{Phys. Rev. Lett.}, 2007, \textbf{99}, 048102\relax
\mciteBstWouldAddEndPuncttrue
\mciteSetBstMidEndSepPunct{\mcitedefaultmidpunct}
{\mcitedefaultendpunct}{\mcitedefaultseppunct}\relax
\EndOfBibitem
\bibitem[Ebbens \emph{et~al.}(2012)Ebbens, Tu, Howse, and
  Golestanian]{ebbens12}
S.~Ebbens, M.-H. Tu, J.~R. Howse and R.~Golestanian, \emph{Phys. Rev. E}, 2012,
  \textbf{85}, 020401\relax
\mciteBstWouldAddEndPuncttrue
\mciteSetBstMidEndSepPunct{\mcitedefaultmidpunct}
{\mcitedefaultendpunct}{\mcitedefaultseppunct}\relax
\EndOfBibitem
\bibitem[Brown and Poon(2014)]{brown14}
A.~T. Brown and W.~C.~K. Poon, \emph{Soft Matter}, 2014, \textbf{10},
  4016--4027\relax
\mciteBstWouldAddEndPuncttrue
\mciteSetBstMidEndSepPunct{\mcitedefaultmidpunct}
{\mcitedefaultendpunct}{\mcitedefaultseppunct}\relax
\EndOfBibitem
\bibitem[Ebbens \emph{et~al.}(2014)Ebbens, Gregory, Dunderdale, Howse, Ibrahim,
  Liverpool, and Golestanian]{ebbens14}
S.~Ebbens, D.~A. Gregory, G.~Dunderdale, J.~R. Howse, Y.~Ibrahim, T.~B.
  Liverpool and R.~Golestanian, \emph{Euro. Phys. Lett.}, 2014, \textbf{106},
  58003\relax
\mciteBstWouldAddEndPuncttrue
\mciteSetBstMidEndSepPunct{\mcitedefaultmidpunct}
{\mcitedefaultendpunct}{\mcitedefaultseppunct}\relax
\EndOfBibitem
\bibitem[Marchetti \emph{et~al.}(2014)Marchetti, Joanny, Ramaswamy, Liverpool,
  Prost, Rao, and Simha]{Joanny2014}
M.~C. Marchetti, J.~F. Joanny, S.~Ramaswamy, T.~B. Liverpool, J.~Prost, M.~Rao
  and R.~A. Simha, \emph{Rev. Mod. Phys.}, 2014, \textbf{85}, 1143\relax
\mciteBstWouldAddEndPuncttrue
\mciteSetBstMidEndSepPunct{\mcitedefaultmidpunct}
{\mcitedefaultendpunct}{\mcitedefaultseppunct}\relax
\EndOfBibitem
\bibitem[Volpe \emph{et~al.}(2011)Volpe, Buttinoni, Vogt, K{\"u}mmerer, and
  Bechinger]{volpe11}
G.~Volpe, I.~Buttinoni, D.~Vogt, H.-J. K{\"u}mmerer and C.~Bechinger,
  \emph{Soft Matter}, 2011, \textbf{7}, 8810\relax
\mciteBstWouldAddEndPuncttrue
\mciteSetBstMidEndSepPunct{\mcitedefaultmidpunct}
{\mcitedefaultendpunct}{\mcitedefaultseppunct}\relax
\EndOfBibitem
\bibitem[Fily and Marchetti(2012)]{fily12}
Y.~Fily and M.~C. Marchetti, \emph{Phys. Rev. Lett.}, 2012, \textbf{108},
  235702\relax
\mciteBstWouldAddEndPuncttrue
\mciteSetBstMidEndSepPunct{\mcitedefaultmidpunct}
{\mcitedefaultendpunct}{\mcitedefaultseppunct}\relax
\EndOfBibitem
\bibitem[Stenhammar \emph{et~al.}(2013)Stenhammar, Tiribocchi, Allen,
  Marenduzzo, and Cates]{stenhammar13}
J.~Stenhammar, A.~Tiribocchi, R.~J. Allen, D.~Marenduzzo and M.~E. Cates,
  \emph{Phys. Rev. Lett.}, 2013, \textbf{111}, 145702\relax
\mciteBstWouldAddEndPuncttrue
\mciteSetBstMidEndSepPunct{\mcitedefaultmidpunct}
{\mcitedefaultendpunct}{\mcitedefaultseppunct}\relax
\EndOfBibitem
\bibitem[Dunkel \emph{et~al.}(2013)Dunkel, Heidenreich, Drescher, Wensink,
  B{\"a}r, and Goldstein]{dunkel13}
J.~Dunkel, S.~Heidenreich, K.~Drescher, H.~H. Wensink, M.~B{\"a}r and R.~E.
  Goldstein, \emph{Phys. Rev. Lett.}, 2013, \textbf{110}, 228102\relax
\mciteBstWouldAddEndPuncttrue
\mciteSetBstMidEndSepPunct{\mcitedefaultmidpunct}
{\mcitedefaultendpunct}{\mcitedefaultseppunct}\relax
\EndOfBibitem
\bibitem[Hatwalne \emph{et~al.}(2004)Hatwalne, Ramaswamy, Rao, and
  Simha]{Simha2004}
Y.~Hatwalne, S.~Ramaswamy, M.~Rao and R.~A. Simha, \emph{Phys. Rev. Lett.},
  2004, \textbf{92}, 118101\relax
\mciteBstWouldAddEndPuncttrue
\mciteSetBstMidEndSepPunct{\mcitedefaultmidpunct}
{\mcitedefaultendpunct}{\mcitedefaultseppunct}\relax
\EndOfBibitem
\bibitem[Z\"ottl and Stark(2014)]{zottl14}
A.~Z\"ottl and H.~Stark, \emph{Phys. Rev. Lett.}, 2014, \textbf{112},
  118101\relax
\mciteBstWouldAddEndPuncttrue
\mciteSetBstMidEndSepPunct{\mcitedefaultmidpunct}
{\mcitedefaultendpunct}{\mcitedefaultseppunct}\relax
\EndOfBibitem
\bibitem[Schwarz-Linek \emph{et~al.}(2015)Schwarz-Linek, Arlt, Jepson, Dawson,
  Vissers, Miroli, Pilizota, Martinez, and Poon]{Jana2015}
J.~Schwarz-Linek, J.~Arlt, A.~Jepson, A.~Dawson, T.~Vissers, D.~Miroli,
  T.~Pilizota, V.~A. Martinez and W.~C.~K. Poon, \emph{Colloids Surf.}, 2015,
  in press, (= arXiv: 1506.04562)\relax
\mciteBstWouldAddEndPuncttrue
\mciteSetBstMidEndSepPunct{\mcitedefaultmidpunct}
{\mcitedefaultendpunct}{\mcitedefaultseppunct}\relax
\EndOfBibitem
\bibitem[Soto and Golestanian(2014)]{soto14}
R.~Soto and R.~Golestanian, \emph{Phys. Rev. Lett.}, 2014, \textbf{112},
  068301\relax
\mciteBstWouldAddEndPuncttrue
\mciteSetBstMidEndSepPunct{\mcitedefaultmidpunct}
{\mcitedefaultendpunct}{\mcitedefaultseppunct}\relax
\EndOfBibitem
\bibitem[Uspal \emph{et~al.}(2015)Uspal, Popescu, Dietrich, and
  Tasinkevych]{uspal15}
W.~E. Uspal, M.~N. Popescu, S.~Dietrich and M.~Tasinkevych, \emph{Soft Matter},
  2015, \textbf{11}, 434--438\relax
\mciteBstWouldAddEndPuncttrue
\mciteSetBstMidEndSepPunct{\mcitedefaultmidpunct}
{\mcitedefaultendpunct}{\mcitedefaultseppunct}\relax
\EndOfBibitem
\bibitem[Bickel \emph{et~al.}(2013)Bickel, Majee, and W{\"u}rger]{bickel13}
T.~Bickel, A.~Majee and A.~W{\"u}rger, \emph{Physical Review E}, 2013,
  \textbf{88}, 012301\relax
\mciteBstWouldAddEndPuncttrue
\mciteSetBstMidEndSepPunct{\mcitedefaultmidpunct}
{\mcitedefaultendpunct}{\mcitedefaultseppunct}\relax
\EndOfBibitem
\bibitem[Ginot \emph{et~al.}(2015)Ginot, Theurkauff, Levis, Ybert, Bocquet,
  Berthier, and Cottin-Bizonne]{ginot15}
F.~Ginot, I.~Theurkauff, D.~Levis, C.~Ybert, L.~Bocquet, L.~Berthier and
  C.~Cottin-Bizonne, \emph{Phys. Rev. X}, 2015, \textbf{5}, 011004\relax
\mciteBstWouldAddEndPuncttrue
\mciteSetBstMidEndSepPunct{\mcitedefaultmidpunct}
{\mcitedefaultendpunct}{\mcitedefaultseppunct}\relax
\EndOfBibitem
\bibitem[Theurkauff \emph{et~al.}(2012)Theurkauff, Cottin-Bizonne, Palacci,
  Ybert, and Bocquet]{theurkauff12}
I.~Theurkauff, C.~Cottin-Bizonne, J.~Palacci, C.~Ybert and L.~Bocquet,
  \emph{Phys. Rev. Lett.}, 2012, \textbf{108}, 268303\relax
\mciteBstWouldAddEndPuncttrue
\mciteSetBstMidEndSepPunct{\mcitedefaultmidpunct}
{\mcitedefaultendpunct}{\mcitedefaultseppunct}\relax
\EndOfBibitem
\bibitem[Liao \emph{et~al.}(2007)Liao, Subramanian, DeLisa, Koch, and
  Wu]{liao07}
Q.~Liao, G.~Subramanian, M.~P. DeLisa, D.~L. Koch and M.~Wu, \emph{Phys.
  Fluids}, 2007, \textbf{19}, 061701\relax
\mciteBstWouldAddEndPuncttrue
\mciteSetBstMidEndSepPunct{\mcitedefaultmidpunct}
{\mcitedefaultendpunct}{\mcitedefaultseppunct}\relax
\EndOfBibitem
\bibitem[Drescher \emph{et~al.}(2010)Drescher, Goldstein, Michel, Polin, and
  Tuval]{drescher10}
K.~Drescher, R.~E. Goldstein, N.~Michel, M.~Polin and I.~Tuval, \emph{Phys.
  Rev. Lett.}, 2010, \textbf{105}, 168101\relax
\mciteBstWouldAddEndPuncttrue
\mciteSetBstMidEndSepPunct{\mcitedefaultmidpunct}
{\mcitedefaultendpunct}{\mcitedefaultseppunct}\relax
\EndOfBibitem
\bibitem[Drescher \emph{et~al.}(2011)Drescher, Dunkel, Cisneros, Ganguly, and
  Goldstein]{drescher11}
K.~Drescher, J.~Dunkel, L.~H. Cisneros, S.~Ganguly and R.~E. Goldstein,
  \emph{Proc. Natl Acad. Sci.}, 2011, \textbf{108}, 10940\relax
\mciteBstWouldAddEndPuncttrue
\mciteSetBstMidEndSepPunct{\mcitedefaultmidpunct}
{\mcitedefaultendpunct}{\mcitedefaultseppunct}\relax
\EndOfBibitem
\bibitem[Guasto \emph{et~al.}(2010)Guasto, Johnson, and Gollub]{guasto10}
J.~S. Guasto, K.~A. Johnson and J.~P. Gollub, \emph{Physical review letters},
  2010, \textbf{105}, 168102\relax
\mciteBstWouldAddEndPuncttrue
\mciteSetBstMidEndSepPunct{\mcitedefaultmidpunct}
{\mcitedefaultendpunct}{\mcitedefaultseppunct}\relax
\EndOfBibitem
\bibitem[Chiang and Velegol(2014)]{chiang14}
T.-Y. Chiang and D.~Velegol, \emph{Langmuir}, 2014, \textbf{30},
  2600--2607\relax
\mciteBstWouldAddEndPuncttrue
\mciteSetBstMidEndSepPunct{\mcitedefaultmidpunct}
{\mcitedefaultendpunct}{\mcitedefaultseppunct}\relax
\EndOfBibitem
\bibitem[Ishimoto and Gaffney(2013)]{ishimoto13}
K.~Ishimoto and E.~A. Gaffney, \emph{Phys. Rev. E}, 2013, \textbf{88},
  062702\relax
\mciteBstWouldAddEndPuncttrue
\mciteSetBstMidEndSepPunct{\mcitedefaultmidpunct}
{\mcitedefaultendpunct}{\mcitedefaultseppunct}\relax
\EndOfBibitem
\bibitem[Li and Tang(2009)]{li09}
G.~Li and J.~X. Tang, \emph{Phys Rev Lett}, 2009, \textbf{103}, 078101\relax
\mciteBstWouldAddEndPuncttrue
\mciteSetBstMidEndSepPunct{\mcitedefaultmidpunct}
{\mcitedefaultendpunct}{\mcitedefaultseppunct}\relax
\EndOfBibitem
\bibitem[Li and Ardekani(2014)]{li14}
G.-J. Li and A.~M. Ardekani, \emph{Phys. Rev. E}, 2014, \textbf{90},
  013010\relax
\mciteBstWouldAddEndPuncttrue
\mciteSetBstMidEndSepPunct{\mcitedefaultmidpunct}
{\mcitedefaultendpunct}{\mcitedefaultseppunct}\relax
\EndOfBibitem
\bibitem[Li \emph{et~al.}(2014)Li, Karimi, and Ardekani]{li14b}
G.-J. Li, A.~Karimi and A.~Ardekani, \emph{Rheologica Acta}, 2014, \textbf{53},
  911--926\relax
\mciteBstWouldAddEndPuncttrue
\mciteSetBstMidEndSepPunct{\mcitedefaultmidpunct}
{\mcitedefaultendpunct}{\mcitedefaultseppunct}\relax
\EndOfBibitem
\bibitem[Takagi \emph{et~al.}(2014)Takagi, Palacci, Braunschweig, Shelley, and
  Zhang]{takagi14}
D.~Takagi, J.~Palacci, A.~B. Braunschweig, M.~J. Shelley and J.~Zhang,
  \emph{Soft Matter}, 2014, \textbf{10}, 1784\relax
\mciteBstWouldAddEndPuncttrue
\mciteSetBstMidEndSepPunct{\mcitedefaultmidpunct}
{\mcitedefaultendpunct}{\mcitedefaultseppunct}\relax
\EndOfBibitem
\bibitem[Vigeant \emph{et~al.}(2002)Vigeant, Ford, Wagner, and Tamm]{vigeant02}
M.~A.-S. Vigeant, R.~M. Ford, M.~Wagner and L.~K. Tamm, \emph{Appl. Environ.
  Microbiol.}, 2002, \textbf{68}, 2794--2801\relax
\mciteBstWouldAddEndPuncttrue
\mciteSetBstMidEndSepPunct{\mcitedefaultmidpunct}
{\mcitedefaultendpunct}{\mcitedefaultseppunct}\relax
\EndOfBibitem
\bibitem[Lauga \emph{et~al.}(2006)Lauga, DiLuzio, Whitesides, and
  Stone]{lauga06}
E.~Lauga, W.~R. DiLuzio, G.~M. Whitesides and H.~A. Stone, \emph{Biophys. J.},
  2006, \textbf{90}, 400--412\relax
\mciteBstWouldAddEndPuncttrue
\mciteSetBstMidEndSepPunct{\mcitedefaultmidpunct}
{\mcitedefaultendpunct}{\mcitedefaultseppunct}\relax
\EndOfBibitem
\bibitem[Shum \emph{et~al.}(2010)Shum, Gaffney, and Smith]{shum10}
H.~Shum, E.~A. Gaffney and D.~J. Smith, \emph{Proc. Royal Soc. A}, 2010,
  \textbf{466}, 1725--1748\relax
\mciteBstWouldAddEndPuncttrue
\mciteSetBstMidEndSepPunct{\mcitedefaultmidpunct}
{\mcitedefaultendpunct}{\mcitedefaultseppunct}\relax
\EndOfBibitem
\bibitem[Campbell and Ebbens(2013)]{campbell13}
A.~I. Campbell and S.~J. Ebbens, \emph{Langmuir}, 2013, \textbf{29},
  14066\relax
\mciteBstWouldAddEndPuncttrue
\mciteSetBstMidEndSepPunct{\mcitedefaultmidpunct}
{\mcitedefaultendpunct}{\mcitedefaultseppunct}\relax
\EndOfBibitem
\bibitem[Crocker and Grier(1996)]{crocker96}
J.~C. Crocker and D.~G. Grier, \emph{J. Colloid Interface Sci.}, 1996,
  \textbf{179}, 298\relax
\mciteBstWouldAddEndPuncttrue
\mciteSetBstMidEndSepPunct{\mcitedefaultmidpunct}
{\mcitedefaultendpunct}{\mcitedefaultseppunct}\relax
\EndOfBibitem
\bibitem[Vladescu \emph{et~al.}(2014)Vladescu, Marsden, Schwarz-Linek,
  Martinez, Arlt, Morozov, Marenduzzo, Cates, and Poon]{vladescu14}
I.~D. Vladescu, E.~J. Marsden, J.~Schwarz-Linek, V.~A. Martinez, J.~Arlt, A.~N.
  Morozov, D.~Marenduzzo, M.~E. Cates and W.~C.~K. Poon, \emph{Phys. Rev.
  Lett.}, 2014, \textbf{113}, 268101\relax
\mciteBstWouldAddEndPuncttrue
\mciteSetBstMidEndSepPunct{\mcitedefaultmidpunct}
{\mcitedefaultendpunct}{\mcitedefaultseppunct}\relax
\EndOfBibitem
\bibitem[H{\"a}nggi \emph{et~al.}(1990)H{\"a}nggi, Talkner, and
  Borkovec]{hanggi90}
P.~H{\"a}nggi, P.~Talkner and M.~Borkovec, \emph{Rev. Mod. Phys.}, 1990,
  \textbf{62}, 251\relax
\mciteBstWouldAddEndPuncttrue
\mciteSetBstMidEndSepPunct{\mcitedefaultmidpunct}
{\mcitedefaultendpunct}{\mcitedefaultseppunct}\relax
\EndOfBibitem
\bibitem[Cox and Brenner(1967)]{cox67}
R.~G. Cox and H.~Brenner, \emph{Chemical Engineering Science}, 1967,
  \textbf{22}, 1753--1777\relax
\mciteBstWouldAddEndPuncttrue
\mciteSetBstMidEndSepPunct{\mcitedefaultmidpunct}
{\mcitedefaultendpunct}{\mcitedefaultseppunct}\relax
\EndOfBibitem
\bibitem[Kotar \emph{et~al.}(2013)Kotar, Debono, Bruot, Box, Phillips, Simpson,
  Hanna, and Cicuta]{kotar13}
J.~Kotar, L.~Debono, N.~Bruot, S.~Box, D.~Phillips, S.~Simpson, S.~Hanna and
  P.~Cicuta, \emph{Phys. Rev. Lett.}, 2013, \textbf{111}, 228103\relax
\mciteBstWouldAddEndPuncttrue
\mciteSetBstMidEndSepPunct{\mcitedefaultmidpunct}
{\mcitedefaultendpunct}{\mcitedefaultseppunct}\relax
\EndOfBibitem
\bibitem[Gu and Li(2000)]{gu00}
Y.~Gu and D.~Li, \emph{Journal of Colloid and Interface Science}, 2000,
  \textbf{226}, 328--339\relax
\mciteBstWouldAddEndPuncttrue
\mciteSetBstMidEndSepPunct{\mcitedefaultmidpunct}
{\mcitedefaultendpunct}{\mcitedefaultseppunct}\relax
\EndOfBibitem
\bibitem[Lameiras \emph{et~al.}(2008)Lameiras, Souza, Melo, Nunes, and
  Braga]{lameiras08}
F.~S. Lameiras, A.~L.~d. Souza, V.~A. R.~d. Melo, E.~H.~M. Nunes and I.~D.
  Braga, \emph{Mater. Res.}, 2008, \textbf{11}, 217--219\relax
\mciteBstWouldAddEndPuncttrue
\mciteSetBstMidEndSepPunct{\mcitedefaultmidpunct}
{\mcitedefaultendpunct}{\mcitedefaultseppunct}\relax
\EndOfBibitem
\bibitem[Leunissen \emph{et~al.}(2005)Leunissen, Christova, Hynninen, Royall,
  Campbell, Imhof, Dijkstra, Van~Roij, and Van~Blaaderen]{leunissen05}
M.~Leunissen, C.~Christova, A.-P. Hynninen, C.~Royall, A.~Campbell, A.~Imhof,
  M.~Dijkstra, R.~Van~Roij and A.~Van~Blaaderen, \emph{Nature}, 2005,
  \textbf{437}, 235--240\relax
\mciteBstWouldAddEndPuncttrue
\mciteSetBstMidEndSepPunct{\mcitedefaultmidpunct}
{\mcitedefaultendpunct}{\mcitedefaultseppunct}\relax
\EndOfBibitem
\bibitem[Spagnolie \emph{et~al.}(2015)Spagnolie, Moreno-Flores, Bartolo, and
  Lauga]{spagnolie15}
S.~E. Spagnolie, G.~R. Moreno-Flores, D.~Bartolo and E.~Lauga, \emph{Soft
  matter}, 2015, \textbf{11}, 3396--3411\relax
\mciteBstWouldAddEndPuncttrue
\mciteSetBstMidEndSepPunct{\mcitedefaultmidpunct}
{\mcitedefaultendpunct}{\mcitedefaultseppunct}\relax
\EndOfBibitem
\bibitem[Turner \emph{et~al.}(2010)Turner, Zhang, Darnton, and Berg]{turner10}
L.~Turner, R.~Zhang, N.~C. Darnton and H.~C. Berg, \emph{J. Bact.}, 2010,
  \textbf{192}, 3259\relax
\mciteBstWouldAddEndPuncttrue
\mciteSetBstMidEndSepPunct{\mcitedefaultmidpunct}
{\mcitedefaultendpunct}{\mcitedefaultseppunct}\relax
\EndOfBibitem
\bibitem[Merlin \emph{et~al.}(2002)Merlin, McAteer, and Masters]{merlin02}
C.~Merlin, S.~McAteer and M.~Masters, \emph{J. Bact.}, 2002, \textbf{184},
  4573\relax
\mciteBstWouldAddEndPuncttrue
\mciteSetBstMidEndSepPunct{\mcitedefaultmidpunct}
{\mcitedefaultendpunct}{\mcitedefaultseppunct}\relax
\EndOfBibitem
\bibitem[Kim and Karrila(2013)]{kim13}
S.~Kim and S.~J. Karrila, \emph{Microhydrodynamics: principles and selected
  applications}, Courier Dover Publications, 2013\relax
\mciteBstWouldAddEndPuncttrue
\mciteSetBstMidEndSepPunct{\mcitedefaultmidpunct}
{\mcitedefaultendpunct}{\mcitedefaultseppunct}\relax
\EndOfBibitem
\bibitem[Miyamura \emph{et~al.}(1981)Miyamura, Iwasaki, and Ishii]{miyamura81}
A.~Miyamura, S.~Iwasaki and T.~Ishii, \emph{Int. J. Multiphase Flow}, 1981,
  \textbf{7}, 41--46\relax
\mciteBstWouldAddEndPuncttrue
\mciteSetBstMidEndSepPunct{\mcitedefaultmidpunct}
{\mcitedefaultendpunct}{\mcitedefaultseppunct}\relax
\EndOfBibitem
\bibitem[Edelstein \emph{et~al.}(2010)Edelstein, Amodaj, Hoover, Vale, and
  Stuurman]{microManager}
A.~Edelstein, N.~Amodaj, K.~Hoover, R.~Vale and N.~Stuurman, in \emph{Computer
  Control of Microscopes Using $\mu$Manager}, John Wiley, 2010\relax
\mciteBstWouldAddEndPuncttrue
\mciteSetBstMidEndSepPunct{\mcitedefaultmidpunct}
{\mcitedefaultendpunct}{\mcitedefaultseppunct}\relax
\EndOfBibitem
\end{mcitethebibliography}
\bibliographystyle{rsc} %the RSC's .bst file
}

\section*{Appendices}
\appendix

\section{Supplementary Video Information \label{videos}}
~

\textbf{SV1} - Janus swimmers moving through a colloidal crystal in 1$\%$ \ce{H2O2}. Epifluorescence, at 3 fps, $50 \um$ scale bar.\\

\textbf{SV2} - High magnification video of a Janus swimmer orbiting a single colloid inside a crystal at 10$\%$ \ce{H2O2}. Epifluorescence (initially brightfield to show location of neighbouring colloids) at 20 fps, $5 \um$ scale bar.\\

\textbf{SV3} - Confocal video of {\it E. coli} bacteria with green stained bodies and red stained flagella swimming inside a colloidal crystal (green). Colloids  touch each other, but only small, polar end caps are visible. Early in the video, an {\it E. coli} with short flagella orbits the colloid marked with a blue circle. 4 fps, $10 \um$ scale bar.\\

\textbf{SV4} - High magnification, edge-on video of a Janus swimmer orbiting a single colloid at the edge of a crystal in 10$\%$ \ce{H2O2}.  Epifluorescence at 20 fps, $5 \um$ scale bar.

\section{Flagella Stained {\it \textbf{E. coli}} \label{stain}}

Construction of the smooth swimming {\it E. coli} strain AB1157 {\it che}Y has been described previously~\cite{vladescu14}. For the current work, the strain was further modified by replacement of the chromosomal copy of the {\it fli}C gene with a modified copy encoding a mutant FliC protein in which the serine amino acid at position 353 is replaced with a cysteine amino acid. Strain HCB1668 is a Tn5 {\it fli}C null derivative of AW405 in which FliC S353C is expressed from the plasmid pBAD33~\cite{turner10}. This plasmid was used as a template to amplify 803 bp of {\it fli}C by PCR. This encompassed the AGT to TGC mutation which was flanked on each side by 400 bp of the {\it fli}C gene. The primers used for amplification were GCAA\underline{CTCGAG}CAATTGAGGGTGTTTATACTGA and GCAA\underline{\underline{GTCGAC}}CCTGGTTAGCTTTTGCCAACA. Restriction sites for \underline{XhoI} and \underline{\underline{SalI}} were included. The PCR product was purified, digested with XhoI and SalI and ligated into the plasmid pTOF24, which had been digested with the same enzymes. The resultant recombinant plasmid pTOF24 fliC was transformed into AB1157 {\it che}Y and used to replace the wild type {\it fli}C allele with the {\it fli}C mutation by plasmid mediated gene replacement using a previously published method~\cite{merlin02}. Correct insertion of the mutation was verified by sequencing.

The resultant strain AB1157 {\it che}Y pHC60 FliC S353C was grown from a single colony in 10 ml Luria-Bertani broth containing 30 $\ugperml$ kanamycin and 5 $\ugperml$ tetracycline overnight at 30 C and 200 rpm. Bacteria were diluted 1:100 into 35 ml tryptone broth containing antibiotics as above and grown for further 4 h. Next, three washes were performed using phosphate motility buffer (6.2 mM \ce{K2HPO4}, 3.8 mM \ce{KH2PO4}, 67 mM NaCl, 0.1 mM EDTA at pH 7.0) and cells concentrated to a total volume of $\sim$3 ml. To perform flagella labelling the protocol of Turner et al.~\cite{turner10} was followed. Briefly, 10 $\ul$ of Alexa Fluor 633 C5 maleimide (1 $\mgperml$ in dimethyl sulfoxide, Molecular Probes) was added to 1 ml of washed bacteria and incubated at room temperature and 100 rpm for 60 min. Three washes were performed as described above and final density was adjusted to optical density 0.3 at 600 nm in motility buffer containing 0.002 wt$\%$ TWEEN 20. 

\section{Geometrical Considerations \label{app: geometry}}

In this section, we give details of how we estimate the gap sizes and inclination angles between the surface of the swimmer, and the static colloid and glass surfaces. 

\begin{figure}[ht!]
\centering
\includegraphics[width=8 cm]{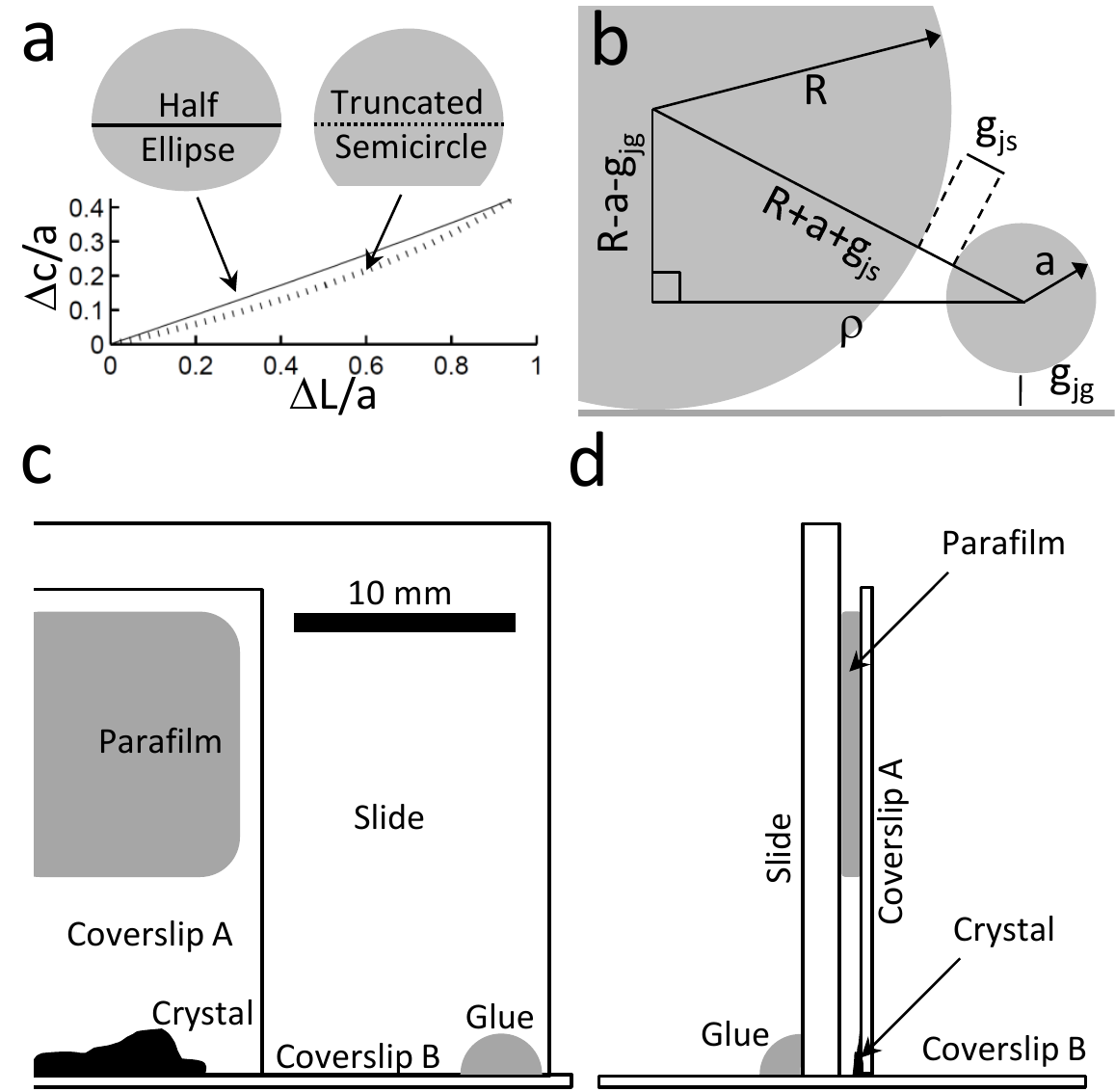}
\caption{a) Results of estimating the offset of a fitted ellipse $\Delta c/a$ from the difference in fitted axis lengths $\Delta L/a$ based on two models shown here and described in the text. b) Side view of a swimmer orbiting a colloid (not to scale) with geometrical construction to determine the size of the gaps $g_{\rm js}$ and $g_{\rm jg}$ between the swimmer and the colloid or plane. c-d) Diagrams of the sample cell for observation along the plane of the coverslip. Observation is from below coverslip B, through the crystal and along the plane of coverslip A. Diagram c) has been cut in half along the left edge of the figure.}
\label{geometry}
\end{figure}

As the swimmer orbits a single colloid, we wish to measure the radius $\rho$ of its orbit, the azimuthal angle of the swimmer around its orbit $\phi$, and the inclination $\beta$ and $\tau$ of the swimmer's orientation away from the tangent to that orbit (see Fig.~\ref{geometry}c). However, since the Janus particle has non-uniform fluorescence intensity, we cannot straightforwardly determine the centre of the particle. We instead measure equivalent parameters ($\rho'$, $\phi'$, $\beta'$) for an ellipse fitted to a thresholded image of the swimmer at each frame, whose centre will be offset from the true centre of the swimmer by some small distance $\Delta c$ along the swimmer's orientation vector. 

The expected shape of the image of the swimmer is not clear, since the Pt coating appears to only partially block out the underlying fluorescence (see supplementary video SV2$^\dag$). We estimate $\Delta c$ from the aspect ratio of the fitted ellipse by performing idential ellipse fitting in MATLAB on two models of the changing thresholded shape of the swimmer, which take the lower half of the image to be either a half-ellipse or a truncated semicircle (Fig.~\ref{geometry}a). 

We plot the relationship between the difference $\Delta L$ in the fitted major and minor axis lengths, and the offset of the centroid $\Delta c$ in these two models, and use the average of these two curves to estimate the experimental value of $\Delta c$. The radius of the Janus swimmers is $a=0.96\pm 0.04~\um$, and, averaging over 17 videos, we find $\Delta L=360\pm 20$ nm, giving $\Delta c=135\pm 30$ nm, where the difference between the two curves in Fig.~\ref{geometry}a has been taken into account in the uncertainty. To lowest order in $\Delta c$, the corrections to $\rho$, $\phi$ and $\beta$ are then given by
\begin{eqnarray*}
	\rho \,=\,\rho'-\Delta c \langle\sin\beta'\rangle\,,
\end{eqnarray*}
\begin{eqnarray}
	\phi \,=\,\phi'-\frac{\Delta c} {\langle \rho' \rangle}\langle\cos\beta'\rangle\,,
\end{eqnarray}
\begin{eqnarray*}
	\beta \,=\,\beta'-\frac{\Delta c} {\langle \rho'\rangle}\langle\cos\beta'\rangle\,.
\end{eqnarray*}
The corrections are approximately 20 nm, and $2^\circ$ respectively, and these have already been applied here and in the body of the article, to give $\langle \rho\rangle=4.56\pm 0.02~\um$ and $\langle\beta\rangle=7\pm2^\circ$.

From the average value of the orbital radius $\langle \rho\rangle$, we calculate the size of the gaps between the swimmer surface and the static colloid $g_{\rm js}$ or the plane glass surface $g_{\rm jg}$. The geometric construction in Fig.~\ref{geometry}b gives the following expression for $g_{\rm js}$ and $g_{\rm jg}$
\begin{eqnarray}
	\rho^2+\left(R-g_{\rm jg}-a\right)^2\,=\,\left(R+g_{\rm js}+a\right)^2\,, \label{geomCon}
\end{eqnarray}
where the radius of the static colloids, $R=5.06\pm 0.02~\um$, and the averages $\langle \ldots \rangle$ have been dropped for convenience. This single equation cannot be used to solve for both $g_{\rm js}$ and $g_{\rm jg}$. However, since both gap sizes must be positive, we can obtain upper bounds on each
\begin{equation}
	\begin{aligned}
		g_{\rm jg}\,<\,\frac{\rho^2-4Ra}{2(R-a)}\,,\\
		g_{\rm js}\,<\,\frac{\rho^2-4Ra}{2(R+a)}\,,
	\end{aligned}
\end{equation}
where we have ignored small terms quadratic in $g_{\rm jg}$, $g_{\rm js}$. Calculating these limits gives $g_{\rm js}<200$ nm and $g_{\rm jg}<300$ nm, taking the largest value within experimental error. If instead, we assume that $g_{\rm jg}=g_{\rm js}=g$, Eq.~\eqref{geomCon} gives
\begin{equation}
	g\,=\,\frac{\rho^2}{4R}-a = 70 \pm 10 \,\mbox{nm}\,. \label{gap}
\end{equation}

For an order-of-magnitude consistency check on this value, we note that the measured translational diffusivity within each orbit, $D_{\rm c}$, is a factor of $\sim3$ smaller than the predicted Stokes-Einstein bulk diffusivity ${D_{\rm SE}=k_BT/(6\pi\eta a)=0.23~\um^2\pers}$ in the bulk. Proximity to walls generally decreases the translational diffusivity of particles~\cite{kim13}. Our geometry is rather complex, and, to our knowledge, has not been treated theoretically. However, since the swimmer is close to two surfaces, and cannot rotate, we may expect a reduction in diffusivity similar to that for a particle moving in the central plane between two parallel plates. Experiments have been performed measuring the drag on spheres moving axially along the centre line of rectangular prisms of aspect ratios 10:1, and these also find a correction factor $\approx 3$ for $g/a=0.1$~\cite{miyamura81}, where $g$ is the gap width. Equation~\ref{gap} gives $g/a\sim 0.07$ in our case, so that the similarity in the diffusivity reduction factors is reassuring.\footnote{Note that the diffusivity measured on plain glass $D_{\rm g}$ is similar to the Stokes-Einstein prediction. However, without a precise measurement of the bulk diffusivity or viscosity, we cannot use this measurement to estimate the swimmer-surface gap in this case.}

To obtain the inclination $\tau$ w.r.t. the glass plane (Fig.~\ref{geometry}c), 10 Janus swimmers orbiting colloids in the crystal were observed along the plane of the coverslip using a custom-built sample chamber, shown in Fig.~\ref{geometry}d-e. A colloidal crystal was formed at the edge of a 22$\times$22 mm$^2$ coverslip (A), as in the main text. Coverslip A was attached with $\sim600~\um$ parafilm to a glass slide previously cut down to 50 mm, so that the edge of coverslip A was flush with the long edge of the slide, with the crystal facing inwards. The slide was then glued onto a 22$\times$50 mm$^2$ coverslip (B), with the crystal lying next to coverslip B. Janus swimmers in 10$\%$ \ce{H2O2} solution were added as usual, and viewed through coverslip B using a 100$\times$ oil immersion objective. Swimmers were recorded orbiting single colloids at the lower edge of the crystal, and images were captured with a CoolSNAP (Photometrics) camera using MicroManager~\cite{microManager} (see SV4$^\dag$). The inclination $\tau=1\pm2^\circ$ of the swimmers w.r.t. coverslip A was determined by fitting ellipses to thresholded images of the swimmers, as above.

\section{Hydrodynamic Interactions \label{hydro}}

In this section, we write down, for a swimmer moving in a circular orbit in free space, the speed variation induced by hydrodynamic interaction with a spherical object outside that orbit. The swimmer is modelled as a stresslet of strength $\alpha$, oriented along a swimming direction $\hat{\mathbf{v}}$. The swimmer is instantaneously located at position $\mathbf{s}$, lying on a circular orbit whose local tangent vector is $\hat{\mathbf{p}}$. A colloid of radius $A$ is located at some arbitrary position $\mathbf{X}$. The displacement vector $\mathbf{l}$ of the swimmer from the static colloid is $\mathbf{l}=\mathbf{s}-\mathbf{X}$, with center-to-centre distance $l=|\mathbf{l}|$. The distance between the centre of the swimmer and the surface of the neighbouring colloid is $h=l-R$. 

We decompose the swimmer's orientation into components perpendicular and parallel to the neighbouring colloid's surface, in order to use the expressions for the advected velocity given in~\cite{spagnolie15}. We therefore define two unit vectors, $\hat{\mathbf{l}}$, which is perpendicular to the colloid surface, and $\hat{\mathbf{k}}$ which is parallel to the colloid surface, and lies in the $\hat{\mathbf{v}},\,\hat{\mathbf{l}}$ plane. These two unit vectors are
\begin{align}
	\hat{\mathbf{l}}&\,=\,\frac{\mathbf{l}}{l}\,,\nonumber\\
	\hat{\mathbf{k}}&\,=\,\frac{\hat{\mathbf{l}}\times(\hat{\mathbf{v}}\times\hat{\mathbf{l}})}{|\hat{\mathbf{v}}\times\hat{\mathbf{l}}|}\,,\label{lkdef}
\end{align}
In this coordinate system
\begin{eqnarray}
	\hat{\mathbf{v}}\,=\, \hat{\mathbf{k}}\cos{\omega}+\hat{\mathbf{l}}\sin{\omega}\,,
\end{eqnarray} 
where $\omega$ is the inclination of the swimmer away from the tangent plane to the colloid's surface ($\sin{\omega}=\hat{\mathbf{v}}\cdot\hat{\mathbf{l}}$). We can define two other angles likewise: $\sin{\psi}=\hat{\mathbf{p}}\cdot\hat{\mathbf{l}}$, and $\cos{\chi}=\hat{\mathbf{p}}\cdot\hat{\mathbf{v}}$. 

The hydrodynamic interactions between a free swimmer, moving originally at speed $u_0$ along direction $\hat{\mathbf{v}}$, and the sphere, would in general result in an additional swimmer velocity $\mathbf{\Updelta u}$, which can be decomposed along $\hat{\mathbf{l}}$ and $\hat{\mathbf{k}}$
\begin{eqnarray}
	\mathbf{\Updelta u}\,=\,u_l(h,\omega,u_0)\hat{\mathbf{l}}+u_k(h,\omega,u_0)\hat{\mathbf{k}}\,.
\end{eqnarray} 
In the present case, however, the particle velocity is constrained to lie on the tangent, $\hat{\mathbf{p}}$, so the observed variation in swimmer speed will be $u'=\hat{\mathbf{p}}\cdot\mathbf{\Updelta u}$, or
\begin{eqnarray}
	u'\,=\, u_l\sin{\psi}+u_k\frac{\cos{\chi}-\sin{\psi}\sin{\omega} }{\cos{\omega}} \,, \label{speedVariation}
\end{eqnarray} 
where, for the velocity components $u_l$ and $u_k$, we can directly use recently derived far-field interaction formulae~\cite{spagnolie15}. Translating into our coordinate system, these are
\begin{align}
	u_l & = \frac{-3R\alpha\left(1-3\sin^2{\omega}\right)\left( R +h\right)}{2h^2\left(2R+h\right)^2}\nonumber\\
	u_k & = \frac{3R^3\alpha\left(2R^2+6Rh+3h^2\right)\sin{(2\omega)}}{4h^2\left(R+h\right)^3\left(2R+h\right)^2}\,. \label{ukul}
\end{align}
and combining Eq.~\eqref{speedVariation}-\ref{ukul} will then give the predicted fractional speed variation $u'/u_0$.

It remains to write down the relevant coordinates. The (fictitious) glass surface is on the $x-y$ plane, with $z$ pointing into the sample, and the origin is at the point of contact between the (fictitious) central colloid and the plane. We take the swimmer to be a small distance $g_{\rm jg}$ above the plane, and orbiting at horizontal distance $\rho$ from the z-axis through the centre of the central colloid ($x=y=0$), and define its position $\mathbf{s}$ in terms of the azimuthal angle $\phi$
\begin{eqnarray}
	\mathbf{s}\,=\,\rowvec{3}{\rho\cos{\phi},}{\rho\sin{\phi},}{a+g_{\rm jg}}\,.
\end{eqnarray}
The neighbouring colloid is fixed at
\begin{eqnarray}
	\mathbf{X}\,=\,\rowvec{3}{2R,}{0,}{R}\,,
\end{eqnarray} 
while the tangent to the circular orbit of the swimmer is
\begin{eqnarray}
	\hat{\mathbf{p}}\,=\,\rowvec{3}{-\sin{\phi},}{\cos{\phi},}{0}\,,
\end{eqnarray} 
and the orientation of the swimmer is
\begin{equation}
	\hat{\mathbf{v}}=\rowvec{3}{-\sin{(\phi-\beta)}\cos{\tau},}{\!\!\cos{(\phi-\beta)}\!\cos{\tau},}{\!\!\sin{\tau}}\!,
\end{equation} 
where $\beta$ is the fixed angle between the tangent to the orbit and the orientation of the swimmer in the $x$-$y$ plane, and $\tau$ is the fixed inclination of the swimmer away from the horizontal plane (Fig.~\ref{geometry}c). This gives $\cos{\chi}=\cos{\beta}\cos{\tau}$.

%The \balance command can be used to balance the columns on the final page if desired. It should be placed anywhere within the first column of the last page.

%\balance

%If notes are included in your references you can change the title from `References' to `Notes and references' using the following command:
%\renewcommand\refname{Notes and references}

\end{document}